\documentclass[journal=jacsat,manuscript=article]{achemso}

\usepackage[version=3]{mhchem} 

\usepackage[utf8]{inputenc}
\usepackage{chemformula} 
\usepackage[T1]{fontenc} 
\usepackage{scalerel,amssymb}
\usepackage{booktabs}
\usepackage{siunitx}
\usepackage{nameref}
\usepackage{hyperref}
\usepackage{subcaption}
\usepackage[labelfont=bf]{caption}

\author{Gihan Panapitiya}
\email{gihan.panapitiya@pnnl.gov}
\author{Michael Girard}
\affiliation[Unknown University]
{Pacific Northwest National Laboratory}
\author{Aaron Hollas}
\author{Vijay Murugesan}
\author{Wei Wang}
\author{Emily Saldanha}
\email{emily.saldanha@pnnl.gov}

\title{Predicting Aqueous Solubility of Organic Molecules Using Deep Learning Models with Varied Molecular Representations}


\begin{document}

\maketitle

\begin{abstract}
Determining the aqueous solubility of molecules is a vital step in many pharmaceutical, environmental, and energy storage applications. Despite efforts made over decades, there are still challenges associated with developing a solubility prediction model with satisfactory accuracy for many of these applications. The goal of this study is to develop a general model capable of predicting the solubility of a broad range of organic molecules. Using the largest currently available solubility dataset, we implement deep learning-based models to predict solubility from molecular structure and explore several different molecular representations including molecular descriptors, simplified molecular-input line-entry system (SMILES) strings, molecular graphs, and three-dimensional (3D) atomic coordinates using four different neural network architectures - fully connected neural networks (FCNNs), recurrent neural networks (RNNs), graph neural networks (GNNs), and SchNet. We find that models using molecular descriptors achieve the best performance, with GNN models also achieving good performance. We perform extensive error analysis to understand the molecular properties that influence model performance, perform feature analysis to understand which information about molecular structure is most valuable for prediction, and perform a transfer learning and data size study to understand the impact of data availability on model performance.
\end{abstract}

\section{Introduction}\label{sec:intro}

Aqueous solubility prediction is one of the key steps in material selection for pharmaceutical, environmental, and renewable energy applications. For example, solubility is a critical physical property for drug development and for methods such as chemical and synthetic route design. In particular, molecular solubility is a key performance driver for redox flow batteries (RFBs) based on organic active materials. These are a promising energy storage technology with potential to address the cost, safety, and functionality needs of the grid-scale energy storage systems forming a critical component of our future electric grid for renewable integration and grid modernization~\cite{chen2017}. The key feature of RFB technology is that the energy-bearing redox-active ions/molecules are dissolved in a supporting liquid electrolyte, which, in the case of aqueous RFBs, is water. Traditional transition metal ions commonly used for RFBs are facing many challenges, such as cost and limited chemical space~\cite{luo2019}, which has led to the search for inexpensive and sustainable organic molecules to support growing grid energy storage needs. Because the solubility of candidate organic molecules dictates their maximum concentration in an electrolyte, and thus the energy density of a RFB system, solubility is a key molecular design factor. The need to quickly screen and explore potential candidate molecules for their expected performance in the RFB, motivates us to develop improved models for solubility prediction that can perform well at the high solubility level (~>0.5 mol/L) required for these technologies.

Solubility prediction has been an intensive research area for many years. Major approaches include the General Solubility Equation~\cite{sanghvi2003estimation}, the Hildebrand
and Hansen solubility parameters~\cite{hildebrand1964solubility,hansen2007hansen}, COSMO-RS~\cite{klamt2002prediction}, and methods leveraging molecular dynamics simulations~\cite{gupta2011prediction,li2017computational}.
Solubility prediction efforts have increasingly turned to the use of statistical and machine learning methods. Early computational solubility prediction efforts based on molecular structure were mainly based on developing regression models to predict solubility using the structural and electronic properties of the molecules as input. For example, regression models were developed which leveraged connectivity indices and a polarizability factor~\cite{Nirmalakhandan88}, structural and atomic charge based properties~\cite{Bodor91}, and molecular fragments~\cite{Klopman92}. 

As high-performance computers and large training datasets became available, artificial neural networks and deep learning grew in popularity.  Advancements in methods and software enabled researchers to apply these techniques to the improvement of quantitative structure–activity relationship models across a broad range of molecular properties. Traditional machine learning methods require pre-calculated "features" as a way of representing molecular structures. Creating high quality features often demands domain expertise and can be very time consuming. Deep learning methods provide a pathway to bypass the feature generation step as they are capable of learning structure-property relationships directly from inputs representing the raw molecular structure. This capability has motivated scientists working on materials property prediction to develop deep learning methods that work as mapping functions which take raw molecular/crystalline structure as the input and physicochemical properties as output. In recent years, these methods have proven to be promising in predicting thermal conductivity, toxicity, lipophilicity, bioactivity, water solubility, protein structure band gap, heat capacity, and scent descriptors, among other properties~\cite{Lusci13, Xiong2020, Torrisi2020, tang2020,wang2019,gladkikh2020,sanchez2019}.  These efforts have explored a range of molecular representations and deep learning modeling architectures, including molecular fingerprints and fully connected neural networks~\cite{Huuskonen1997, Huuskonen2000, livingstone_2001, Ma2015, Korotcov2017, Tong2020}, simplified molecular-input line-entry system (SMILES) strings and recurrent neural networks~\cite{hirohara2018,arus-pous2019,Olivecrona2017}, molecular graphs and graph neural networks~\cite{Duvenaud_2015, xie2018, coley2019, Lusci13, Xiong2020, tang2020}, and spatially-aware architectures such as SchNet~\cite{Schutt2017,schtt2017schnet}. 

These types of techniques have also been applied to the problem of solubility prediction. The most often applied graph-based neural network techniques include DAG Recursive Neural Networks~\cite{Lusci13}, graph convolutional networks~\cite{coley2017}, message passing neural networks (MPNN)~\cite{withnall2020}, and MPNN models with self-attention~\cite{tang2020}. Other efforts have explored alternative architectures such as \citet{cui2020} who compare the performance of shallow neural networks with deeper ResNet-like networks for solubility prediction. These efforts generally rely on small datasets, ranging from 100 to 1,297 molecules, with the exception of \citet{cui2020}, which leverage a dataset with around 10,000 molecules.

Despite these developments, the prediction of solubility remains challenging. Several of the major challenges for this task include the complexity of the solvation process, the existence of measurement noise and data quality issues, the diversity and scale of the molecular structure space, and the broad range of solubility values.  These values span many orders of magnitude. Many of the described challenges and limitations are driven by the limited size of available datasets, which do not have the needed diversity or capacity for models to learn the complex relationships between structure and solubility. Another direction for addressing these challenges is through the development of improved molecular representations and the application of models with the capacity to learn complex structure-property relationships.   

In this work, we explore the predictive capacity of different commonly used molecular representation approaches and deep learning model variants on the largest and most diverse collection of organic solubility measurements to date. We make several key contributions. First, in contrast to previous efforts, we perform a comparison across all commonly used representations and modeling approaches on the same dataset to determine which are best suited to extract the underlying structure-property relationships and demonstrate that feed-forward networks leveraging molecular descriptors outperform other approaches. While it is challenging to make a direct comparison with previous efforts, due to differences in the datasets, we find that the combination of our models and training dataset lead to equivalent or improved performance on almost all previously used solubility prediction datasets, demonstrating the impact of large training sets on model generalizability. 
Secondly, we perform detailed exploration of the errors made by the resulting models to understand the types of molecular structures for which prediction is successful, and the types for which it is more challenging. We identify the prediction of solubility within groups of isomers as a key challenge for future development.  Finally, we demonstrate the impact of dataset size on the predictive capabilities of the model through a transfer learning evaluation and an exploration of performance on smaller data subsamples. We find that doubling the data size is associated with a reduction in RMSE of 0.06 orders of magnitude and that leveraging transfer learning provides a performance boost for models leveraging raw molecular structure as inputs.

\section{Data}
\label{sec:data}

\begin{figure*}[t]
    \centering
    \includegraphics[width=1\textwidth]{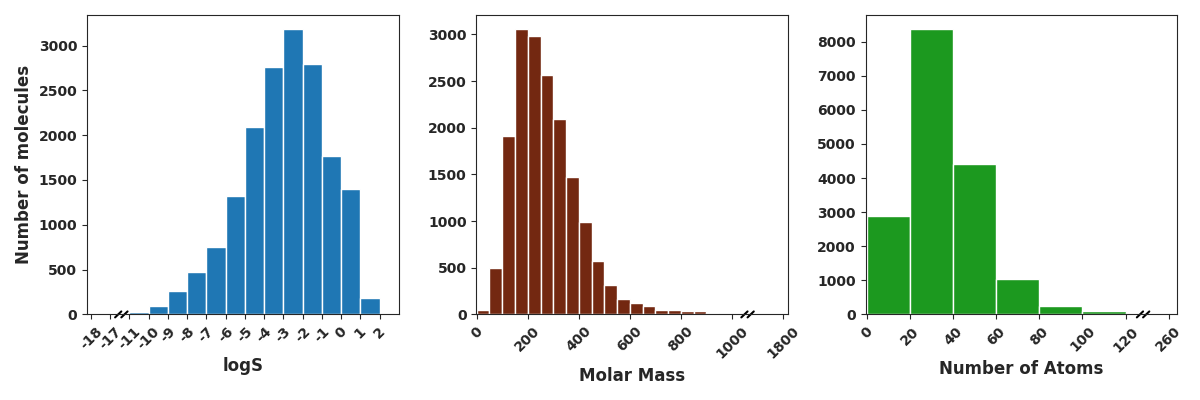}
    \caption{Distributions of log solubility, molar mass, and number of atoms for molecules in our dataset.}
    \label{fig:dist_logS}
\end{figure*}

In order to train our deep learning models we leverage a large dataset compiled by Gao et al. \cite{2021Gao} containing data for 11,868 molecules collected from various data sources (including OChem~\cite{sushko_online_2011}, Beilstein~\cite{Reaxyz}, and Aquasol~\cite{sorkun_2019}). We also use data made available by Cui et al.\cite{cui2020} and  a commercial data set obtained from Reaxys~\cite{Reaxyz}. The combined dataset consists of 17,149 molecules with sizes ranging from 1 to 273 atoms and with molecular masses ranging from 16 to 1,819.
The measured solubilities of these molecules range from $3.4\times10^{-18}$ to 45.5 mol/L. The distributions of log solubility values, molecular mass, and number of atoms are shown in Figure~\ref{fig:dist_logS}. Throughout this manuscript, logS stands for base 10 logarithm value of solubility S, which is in the units of mol/L, where L stands for the volume of the solvent in liters.



In order to study the relationship between solubility and molecular properties as well as to develop features for input to the models, we generate several different sets of features derived from the molecular structure -- two-dimensional (2D) molecular features, three-dimensional (3D) molecular features, functional group features, and DFT-based quantum descriptor features. First, we employed 2D molecular descriptors as implemented in the Mordred package~\cite{Moriwaki_2018}. In total this package can generate 1613 descriptors derived from 2D molecular structures. However, the descriptor generation failed for some molecules in our dataset, and we therefore relied on 743 features which could be successfully generated for all the molecules (these are listed in Tables S1 and S2). This set of features will be referred to as 2D descriptors in the remainder of the text.  

Additionally, we calculated a set of features describing the 3D structure of the molecules (which we will refer to as 3D descriptors). Atomic coordinates for these calculations were generated using the Pybel package~\cite{oboyle_pybel_2008}. The coordinates are optimized using MMFF94 force fields with 550 optimization steps. There were 36 molecules for which coordinate generation failed, which we dropped from the dataset. Using the approximated coordinates, we calculated counts of atoms within six concentric layers around the centroid of the molecule as described in \citet{Panapitiya2018} to be used as features. Another set of features that contain information about distribution of atoms has been proposed by \citet{Ballester2007}. To calculate these features, the distances to all the atoms with respect to three locations in the molecule (centroid, closest atom to the centroid, farthest atom to the centroid) are calculated. Next, we calculate the statistical moments of the atomic distance distributions from order 1 to 10. These features encode information about the shape of the molecule. We also calculated the volume enclosed by all the atoms in a molecule using the ConvexHull function implemented in the Scipy package~\cite{Panapitiya2018, scipy}. In total there are 37 resulting 3D descriptors.



In addition to the molecular descriptor features, we included counts of  molecular fragments and functional groups present in the molecules. First, we identified a set of fragments to use as features.
We used RDKit~\cite{RDKit} to identify molecular fragments attached to benzene-like structures (hexagonal ring with 6 atoms) in our dataset.
From the resulting fragments, we selected the 52 most common fragments in addition to seven other functional groups commonly found in chemical compounds. These 59 fragments are shown in Figure S1. In total, there are 839 molecular descriptors used as features.


\begin{table}[t]
\centering
\footnotesize
\begin{tabular}{lrrrrr}
\toprule
        \textbf{Dataset} &  \textbf{N} & \textbf{logS} &    \textbf{A} & \textbf{AA} &   \textbf{R} \\
\midrule
           Ours & 17,149 & -17.5 - 1.7 &  1 - 273 &   0 - 64 &  0 - 33 \\
           \midrule
        Delaney~\cite{delaney2004} & 1,144 & -11.6 - 1.6 &  4 - 119 &   0 - 28 &   0 - 8 \\
          Tang~\cite{tang2020} & 4,200 & -11.6 - 1.6 &   5 - 94 &   0 - 23 &   0 - 7 \\
            Cui~\cite{cui2020} & 10,166 & -18.2 - 1.7 &  1 - 216 &   0 - 60 &  0 - 16 \\
        Boobier~\cite{boobier_2017} & 100  &    -8 - 1 &  10 - 67 &   0 - 20 &   0 - 7 \\
      Huuskonen~\cite{huuskonen2000_dataset} & 1,297  &   -11 - 1 &   5 - 94 &   0 - 23 &   0 - 7 \\
 Sol. Chall.~\cite{Llinas2008} &  132 &   -7 - -1 &  13 - 76 &   0 - 19 &   1 - 5 \\
\bottomrule
\end{tabular}

\caption{Comparison of the diversity of different datasets, showing the range of values observed in the datasets. N, logS, A, AA, and R refer to the number of molecules, log solubility (mol/L), number of atoms, number of aromatic atoms, and number of rings respectively.}
\label{dataset_complexity}
\end{table}

Finally, in order to asses the impact of features derived from Density Functional Theory (DFT), we leveraged a set of quantum descriptors, including the solvation energy (kcal/mol), molecular volume (Ang$^3$),
molecular surface area (Ang$^2$), dipole moment (Debye),
dipole moment/volume (Debye/A$^3$) and quadrupole moments as calculated using the NWChem package~\cite{nwchem}. Due to the high computational resources it takes to optimize large molecular structures using DFT quantum descriptors, only 7764 molecules containing at most 83 atoms have been used. Therefore, in our primary analysis we exclude these features but perform a study of their impact on the models in Section~\nameref{sec:featureanalysis}.




In order to compare the performance of our models with the results of previous efforts, we perform an evaluation using six previously existing datasets, including  \citet{delaney2004}, \citet{huuskonen2000_dataset}, \citet{boobier_2017}, \citet{tang2020}, \citet{Llinas2008}, and \citet{cui2020}. A summary of different properties of these datasets are given in Table~\ref{dataset_complexity}, Table S3, and Figure S2. Except for the Cui dataset, the others consist of small molecules containing at most eight rings. While the Cui dataset does contain complex molecules, our dataset introduces even further diversity. Because the datasets contain duplicate entries with potentially differing solubilities, for the purposes of our analysis we treat duplicate entries across these datasets according to a method similar to what is used in \citet{sorkun_2019} (described in detail in the supporting information).

To support the prediction of solubility we also explore the use of transfer learning by leveraging large molecular datasets (QM9 and PC9), which do not include solubility labels, but do contain significantly more molecules than our solubility dataset. The QM9 dataset contains 133,885 small molecules with sizes up to nine atoms and composed of only H, C, N, O, and F atoms~\cite{ramakrishnan2014quantum}. For each molecule, the dataset contains 17 energetic, thermodynamic, and electronic properties along with the SMILES structure corresponding to B3LYP relaxation~\cite{ramakrishnan2014quantum}.
The PC9 dataset contains 99,234 unique molecules that are equivalent to those in QM9 in terms of the atomic composition and the maximum number of atoms, but the dataset is designed to improve upon the chemical diversity in comparison with QM9~\cite{glavatskikh2019}. 

\section{Solubility Prediction}

We aim to to develop deep learning models that can infer the solubility of a molecule by exploiting the patterns that exist between structural molecular properties and measured molecular solubility. We include an exploration of such patterns in our dataset in the supporting information. In order to train models that can automatically recognize such patterns, there are various ways of representing a molecule for computational purposes. Of these, representing a molecule as a vector of structural/electro-chemical features, as a SMILES string, as a molecular graph, and as a set of 3D atomic coordinates are widely used methods. We use these four representations to explore which representations are best suited toward high-accuracy solubility prediction and apply several different deep learning architectures that are well-suited to each data format.

The first representational approach relies on a large suite of molecular descriptors which quantify the structural and electro-chemical properties of the molecule. We leverage a fully connected neural network to predict the solubility, given this set of features. The feature set we use includes the 2D descriptors, 3D descriptors, and fragment counts. Before training the models, the features in the training, validation, and test sets were scaled to zero mean and unit variance using transformation parameters based on the training set. We refer to this model as the molecular descriptor model (MDM).

Our second model is based on using the SMILES string representation of each molecule as input to a character-level long short-term memory (LSTM) neural network~\cite{Hochreiter1997}, which is designed to process sequential data such as the character sequences that comprise SMILES strings. We refer to this model as the SMILES model.

Our third model relies on a molecular graph representation, where the atoms and bonds become nodes and edges of a graph respectively, and a Graph Convolutional Network (GCN), which consists of graph convolutional and edge convolutional layers. Each node is initially assigned with a set of features. For this work we used the features defined in the ``atom\_features'' function of the DeepChem library~\cite{deepchem} which include  atomic symbol, degree, implicit valence, total number of hydrogen atoms, and hybridization of the atom as a one-hot encoded vector, whether the atom is aromatic or not as a boolean feature, and the formal charge of the atom (refer to supporting information for more details). The graph neural network then learns to update the node and edges features through an iterative process called message passing.  We refer to this model as the graph neural networks (GNN) model.

\begin{figure*}[t]
    \centering
    \includegraphics[width=1\textwidth]{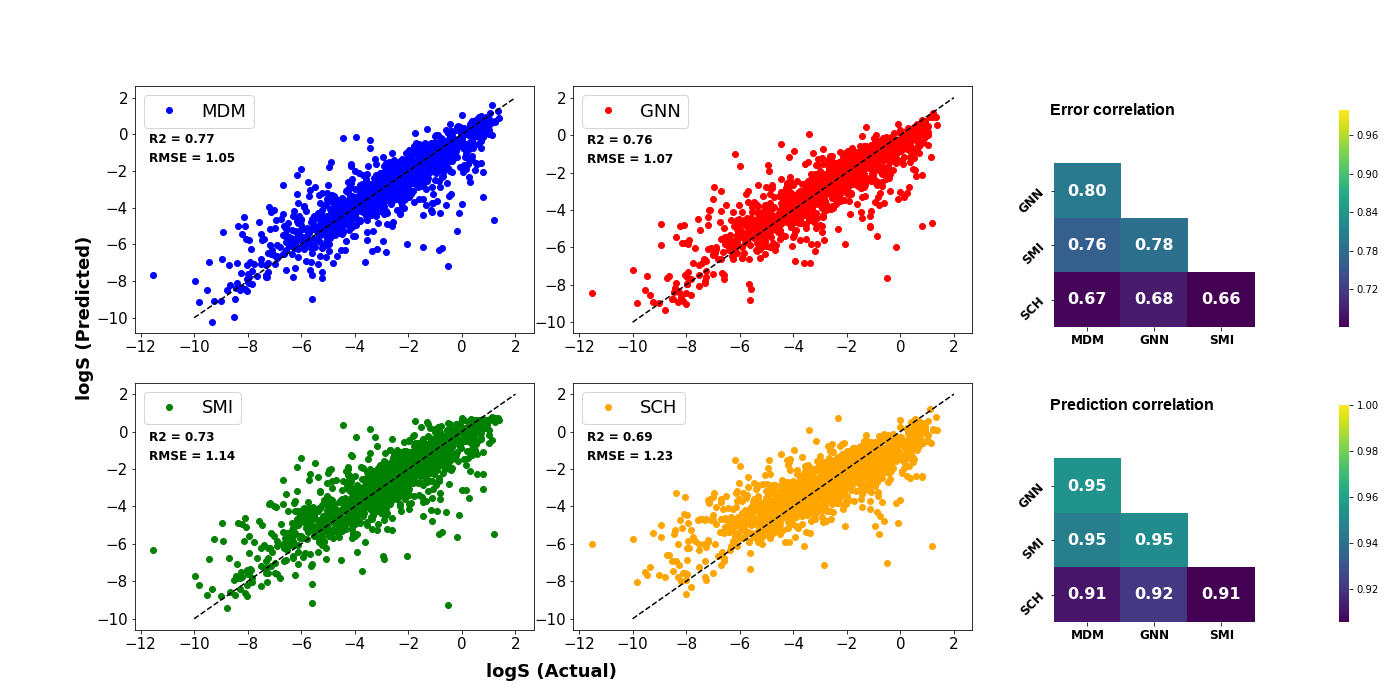}
    
    \caption{Left: Scatter plots of predicted versus actual log solubilities obtained by four models considered in this study. Right: Pearson correlation of errors (top) and predictions (bottom) between different pairs of models.}
    \label{results_scatter}
\end{figure*}%

\begin{table}[t]
\begin{center}
\begin{tabular}{@{}lcccc@{}}    \toprule
\textbf{Model}          & \textbf{R\textsuperscript{2}} & \textbf{Spearman} & \textbf{RMSE} & \textbf{MAE} \\
& &  & \textbf{(log M)} & \textbf{(log M)} \\
\midrule
 MDM         & \textbf{0.7719}  & \textbf{0.8787} & \textbf{1.0513} & \textbf{0.6887}	\\ 
 GNN           & 0.7628  & 0.8708  & 1.0722 & 0.7256 \\ 
 SMILES        & 0.7337 & 0.8603  & 1.1360 & 0.7609 \\ 
 SCHNET        & 0.6883  & 0.8337 & 1.2291 & 0.88924\\ 
 \hline
\end{tabular} 
\end{center}
\caption{Evaluation results for the four models on the test set}
\label{tbl:results}
\end{table}

Finally, we apply a model designed to learn from the full 3D atomic coordinate representation of the molecules called SchNet, originally developed to predict molecular energy and interatomic forces~\cite{schtt2017schnet}.  The SchNet architecture is built upon three types of sub-networks: atom-wise layers, interaction layers, and continuous filter-networks, which learn atom-level representations based on the observed distances between atoms.  We refer to this model as the SchNet model.


\subsection{Optimization}

For the purposes of model development and training, we split our full dataset into three components for training, validation, and testing. Prior to splitting, the solubility values were binned into 6 folds as shown in Figure S10. Next, 85\%, 7.5\%, and 7.5\% of the data were chosen using stratified sampling from the bins for the training, validation, and testing splits respectively. This procedure ensures that high and low solubility molecules are sampled into each of the three splits. Hyperparameter tuning was carried out using the \textit{hyperopt} python package~\cite{hyperopt}. Due to the different training times required by the different models, we were able to perform a larger search of the hyperparameter space for some of the models. For the MDM and GNN models we considered 1,000 different unique combinations, whereas for the SchNet and SMILES models, only 50 and 20 combinations respectively were evaluated. Details on the tuned hyperparameters, their explored ranges, and the final selected parameter values can be found in the supporting information.


\begin{figure}[!h]
    \centering
    \includegraphics[width=.5\textwidth]{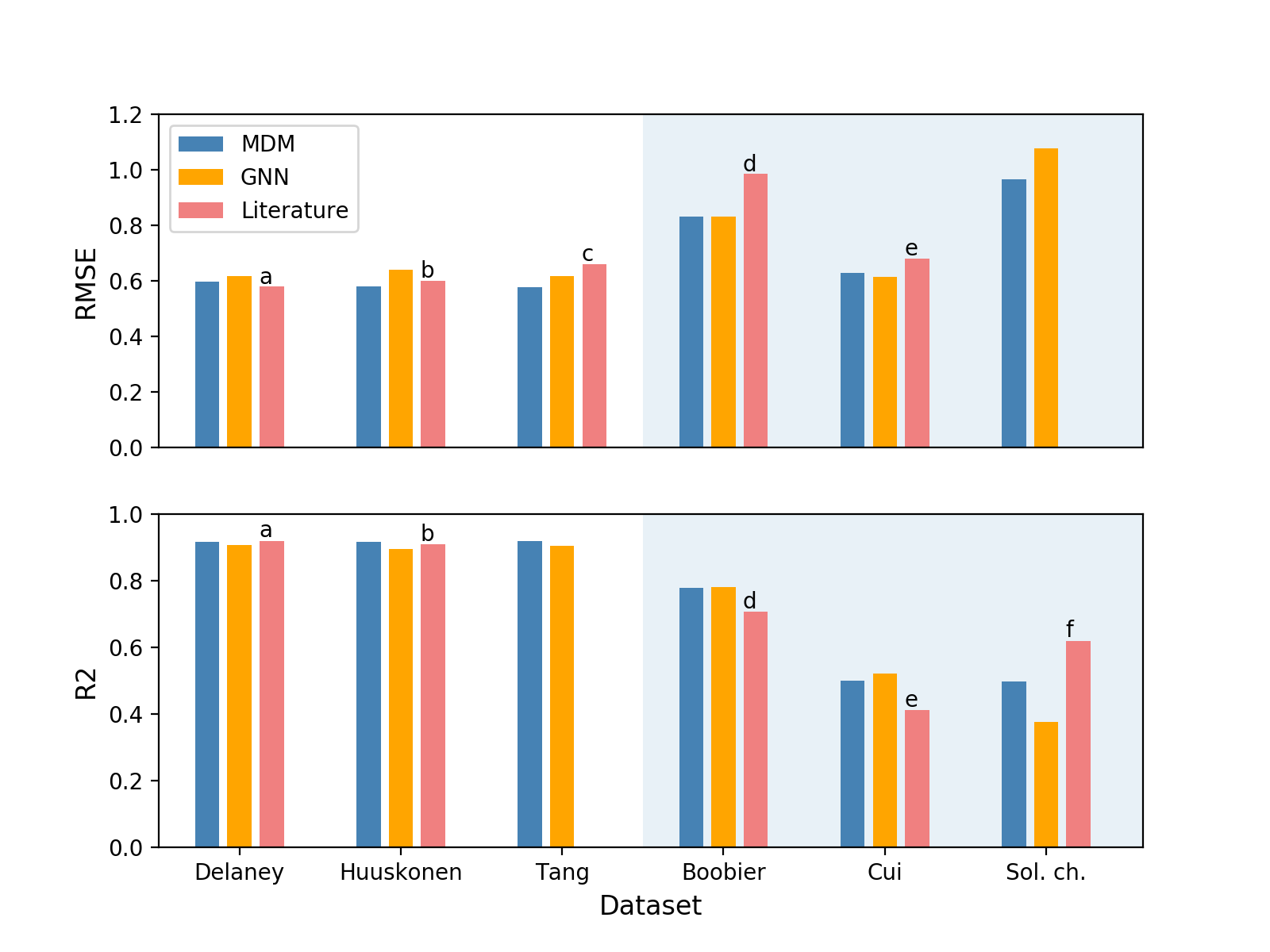}
    \caption{RMSE (top) and $R^2$ (bottom) values of predictions obtained  by MDM and GNN for different datasets using different dataset splitting methods. The datasets on a white background were evaluated using a fixed train-test split and those on a blue background were evaluated using cross-validation. a,b:Lusci et. al.\cite{Lusci13}, c:Tang et. al.\cite{tang2020}, d:Boobier et. al. \cite{boobier_2017}, e: Cui et. al. \cite{cui2020}, f:Hopfinger et. al. \cite{Hopfinger2009} }
    \label{ext_results}
\end{figure}%

\section{Results and Discussion}

We evaluate the performance of each of our representation and modeling approaches using two error metrics, root mean squared error (RMSE) and mean absolute error (MAE) and two correlational metrics, $R^2$ and Spearman correlation. The error metrics allow us to evaluate the mean levels of error observed in the model predictions, while the correlational metrics allow us to observe if the models perform well at ranking the molecules in terms of solubility, even if the exact predictions are not correct. The performance results for each of the models are given in Table \ref{tbl:results} and the predicted versus actual solubility values for all four models are shown in Figure~\ref{results_scatter} (left). We find the best performance is achieved by the MDM model, showing that the models which leverage raw structural information alone are not able to outperform the predictions using pre-derived molecular features on this predictive task.

Of the three models that rely on raw molecular structure information, we find the GNN model achieves the highest performance, almost equaling the performance by the molecular feature model. This shows that GNNs have the capability to learn almost all the information embedded in the molecular features, using only a relatively small number of atomic properties.

We also study the strengths and weakness of the different representations and modeling approaches by observing whether the different models make similar errors. In Figure~\ref{results_scatter} (right), we show the correlation in the predictions and errors for each pair of the models. The high correlation values of the predictions (> 0.9) and errors (> 0.65) show that although the models are using different features and representations of the molecules, they are making very similar predictions. This indicates that the molecules which are easy and hard to predict are largely held in common across the different models, rather than different models excelling for different groups of molecules.

\begin{figure*}[t]
    \centering
    \includegraphics[width=0.90\textwidth]{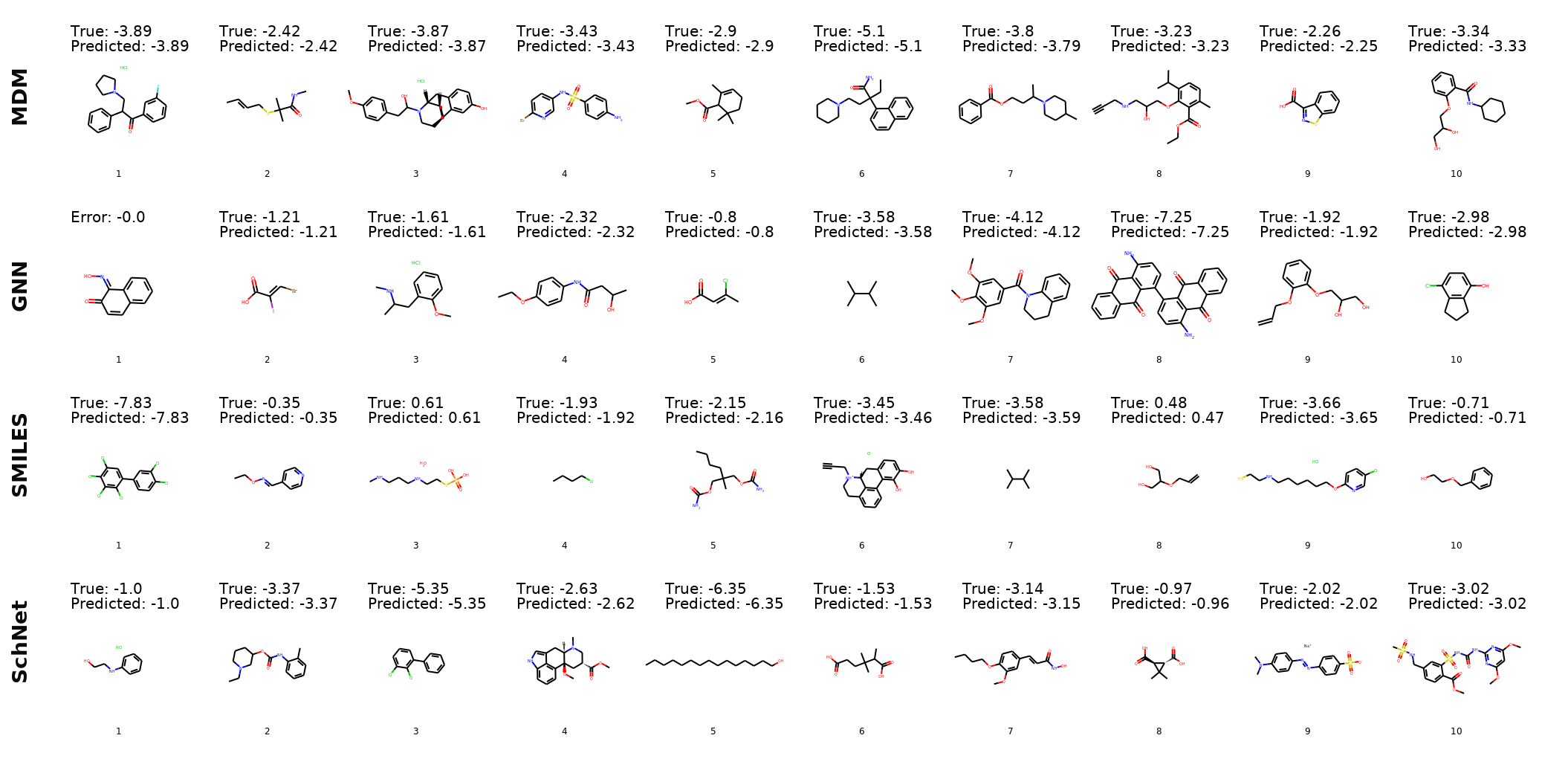}
    \caption{Ten lowest error molecules for each model. For molecules from the commercial database Reaxys, we list error values (true-predicted) rather than providing the true solubility measurement.}

    \label{best_predicted_mols}
\end{figure*}

\begin{figure*}[t]
    \centering
        \includegraphics[width=0.95\textwidth]{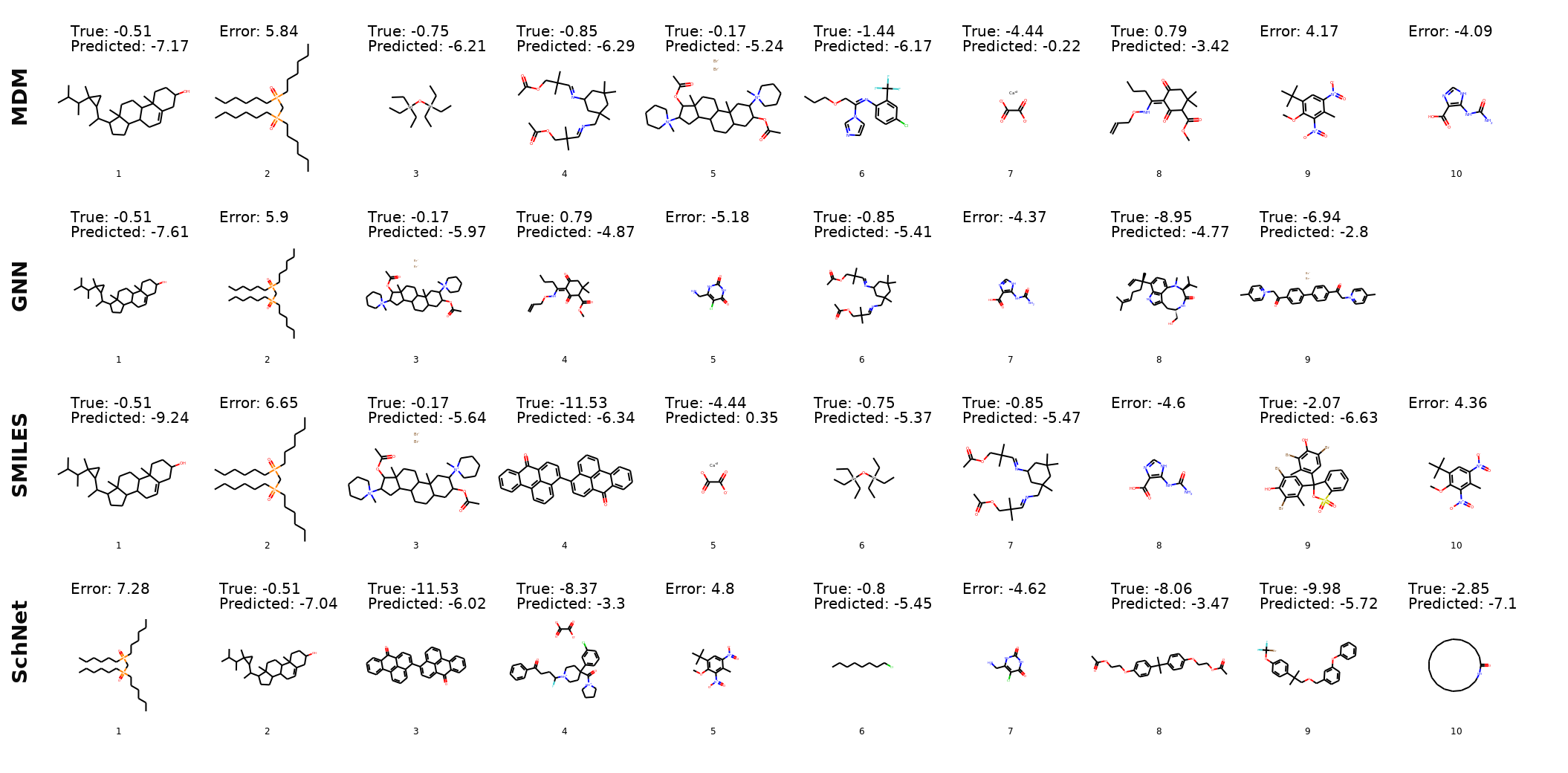}
        
        \caption{Highest error molecules with absolute errors greater than 4 logS. For molecules from the commercial database Reaxys, we list error values (true-predicted) rather than providing the true solubility measurement.}
        \label{worst_predicted_mols}
\end{figure*}

\subsection{Comparison with Previous Results}

To validate the predictive ability of our models, we compared the performance of our modeling approaches with the results obtained in previous solubility prediction studies using six different datasets. These comparison efforts are complicated by the use of differing datasets across many different previous studies, by the fact that previous efforts largely used significantly smaller datasets, and by the overlap of the molecules across the different datasets.  In this comparison, we are aiming to evaluate the impact of both the modeling approach as well as the use of a large and diverse training set of solubility values.

The previous studies used two different strategies for model validation - a fixed test/train split approach and a cross-validation approach where performance is averaged across multiple random splits.  For comparison purposes, we replicate the evaluation approach used by each paper. When the external datasets consist of separate train and test sets, we leverage their training set in combination with ours and test the resulting model performance on the external test set.
For external datasets where the previous authors did not provide separate train/test sets, we used ten-fold cross validation to obtain test results for external datasets. The folds were generated by randomly splitting the external data in ten portions and adding nine of the portions to our training data and using the remaining split as the test set. The final results were calculated by cycling through all ten folds as the test set and averaging the results. We do not perform any new hyperparameter tuning for these models but rely on the parameters determined by optimizing on our dataset alone. 

The resulting model accuracies for six external datasets are given in Figure \ref{ext_results}. We can see that, except for the solubility challenge dataset, the accuracies obtained for other datasets are similar to or better than previous results. In particular, for the three datasets which appear the easiest (Delaney, Huuskonen, and Tang), with low RMSE and high $R^2$ values already previously achieved, our models roughly equal the previously existing performance. This could indicate that there is limited room for predictive performance improvement on these simpler datasets. In contrast, we find that we achieve significant performance improvement for the more challenging Boobier and Cui datasets which have previous $R^2$ results of only 0.71 and 0.42 respectively. These results indicate the potential of a large, diverse dataset in combination with highly expressive deep learning models to learn generalizable structure-property patterns applicable across many different datasets.

The solubility challenge dataset has proven to be the most difficult for our models. The organizers of the solubility challenge competition reported that the solubilities of probenecid, diflunisal, indomethacin, terfenadine, dipyridamole, and folic acid were the least accurately predicted by the competitors (percentages of correct predictions received for these molecules are 2\%, 3\%, 4\%, 6.1\% 12.1\% and 19.2\% respectively)~\cite{Hopfinger2009}. Consistently, our MDM model has also made its worst predictions for these molecules. Except for indomethacin, these molecules are very insoluble in water~\cite{Hopfinger2009}, indicating that machine learning models generally find it difficult to accurately predict low solubilities. This observation agrees with our results shown in Figures~\ref{err_by_range} and S14.

\section{Error Analysis}

Next we perform detailed analysis of the errors made by the models to understand the factors leading to improved and reduced predictive performance. We perform several different analyses of the errors, including manual examination of easy and difficult molecules and performance comparison on molecules of different types.

\subsection{Qualitative Examination}

First, we observe the molecules for which the models have exceptionally low or high error values. Figure \ref{best_predicted_mols} and Figure \ref{worst_predicted_mols} show the top ten molecules with lowest and highest error for each model. We find that the low error predictions of the MDM model are for molecules with log solubilities in the range of -2.26 to -5.1, showing that the greater data availability for this range of solubilities may improve predictions. While there are no common molecules among the low error instances across all four models, we do observe significant overlap in the molecules that proved most difficult for the different models.

By examining the set of high error molecules, we can identify several potential data labeling issues in the dataset. For example, we find that the original reference solubility for molecule 4 (Figure \ref{worst_predicted_mols}) from the high MDM errors is actually the solubility of the decomposed aldehyde product rather than the solubility of the full molecule. For molecule 6 from the high MDM errors, there are two values that exist in the literature, logS = -1.44~\cite{Shiu1990}, which is the value in the current database, and logS = -4.54~\cite{efsa_triflumizole}, which is in better agreement with the model prediction.


When collecting measurement data from multiple online sources to compile a large database, the existence of some level of noise and errors in the data cannot easily be avoided. The process of manual validation of measurements is time consuming and would not be tractable to perform on a database with 17K molecules. The qualitative examination performed here shows that errors made by the predictive models can be used as a signal to identify potential issues arising in the data, informing improvements to future versions of the database. By showing that low performance on some of these molecules can be attributed to data issues rather than true model errors, we also increase confidence in the predictive capabilities of the models.




\begin{figure}[t]
    \centering
        \includegraphics[width=.45\textwidth]{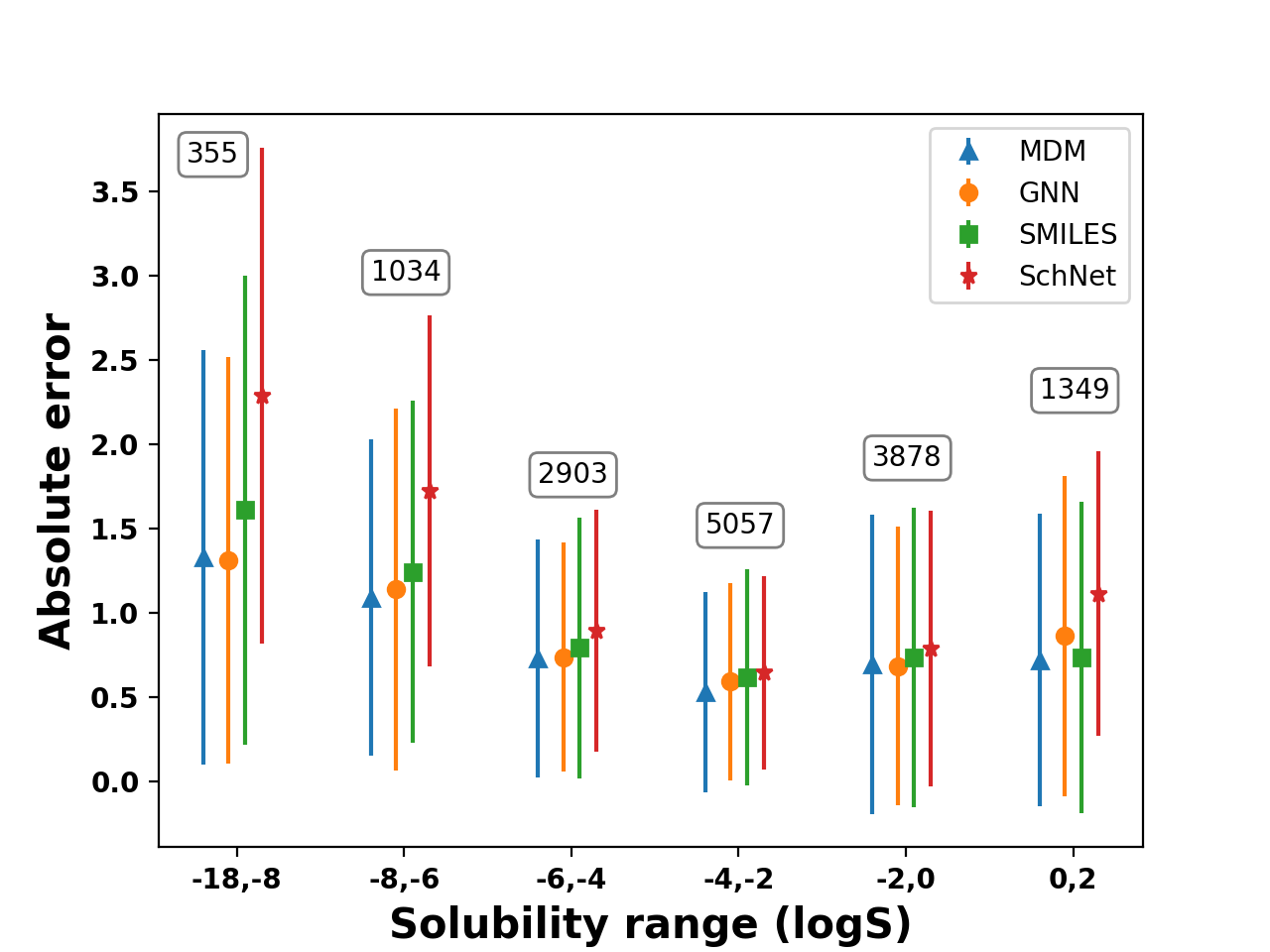}
        
        \caption{Test set error by solubility range (in logS) for the four models with the number of test molecules in each bin annotated on the plot, showing the mean and standard deviation per bin.}
        \label{err_by_range}
\end{figure}
 
\subsection{Errors By Solubility}
Next, we observe whether there is any relationship between model error and measured solubility of the molecules. We binned the molecules into solubility ranges and calculated mean and standard deviation of model errors on the test set in each bin as shown in Figure \ref{err_by_range}. The corresponding number of training data points for each range are also shown. We find that generally solubility ranges with more data are easier to predict, showing the impact of training data size on model performance. We also find that the models generally have worse errors for low solubility molecules, with higher solubility molecules being easier to predict. However, we should also keep in mind that the predictive task is performed on log solubility, which means that an absolute error of two orders of magnitude represents a much smaller actual error for low solubility bins than it does for high solubility bins.



\subsection{Errors by Molecule Type}
Next we aim to determine whether certain types of molecules are more challenging for the model to predict. We select several subsets of our dataset by molecule type, such as chiral molecules and inorganic molecules and analyze the model performance for these subsets. The results of this analysis are shown in Table~\ref{tbl:err-by-type}. It is interesting to note that chiral compounds can be predicted with better than average accuracies given that the input molecular representations may be less sensitive to stereochemistry. We also find that molecules in our dataset that fall into groups of isomers are relatively easy to predict. However, we will show in Section~\nameref{sec:groups} that it is difficult for models to distinguish the solubility of molecules \textit{within} individual groups of isomers. Even though there are 2,580 salts and organo-metallic compounds in the training set, the model has found it difficult to learn a generalized mapping function for this group of compounds as we see reduced performance of this group compared with chiral molecules and isomers. It should also be noted that 99\% of molecules in this subset are composed of multiple fragments. Finally, we note that molecules which do not fall in any of these predetermined groups are the hardest subset to predict.



\begin{table}[t]
\caption{Test set errors by molecule type. $N$ is the number of molecules of each type in the test set.}
\begin{center}

\begin{tabular}{lr|rr|rr}
\toprule
            &   &     \multicolumn{2}{c|}{\textbf{MDM}} &     \multicolumn{2}{c}{\textbf{GNN}} \\

         \textbf{Group} &    \textbf{N} &  \textbf{R\textsuperscript{2}} &  \textbf{RMSE} &  \textbf{R\textsuperscript{2}} &  \textbf{RMSE} \\
\midrule
All & & 0.77 & 1.05 & 0.76 & 1.07 \\
\midrule
        Chiral &  142 &    0.85 &      0.92 &    0.81 &      1.03 \\
 Salts \& Org.M &  230 &    0.77 &      1.08 &    0.76 &      1.11 \\
       Isomers &   90 &     0.90 &      0.76 &    0.87 &      0.89 \\
     All other &  857 &    0.74 &      1.08 &    0.74 &      1.08 \\
\bottomrule
\end{tabular}

\end{center}

\label{tbl:err-by-type}
\end{table}

\begin{figure}[t]
    \centering
    
    \includegraphics[width=.5\textwidth]{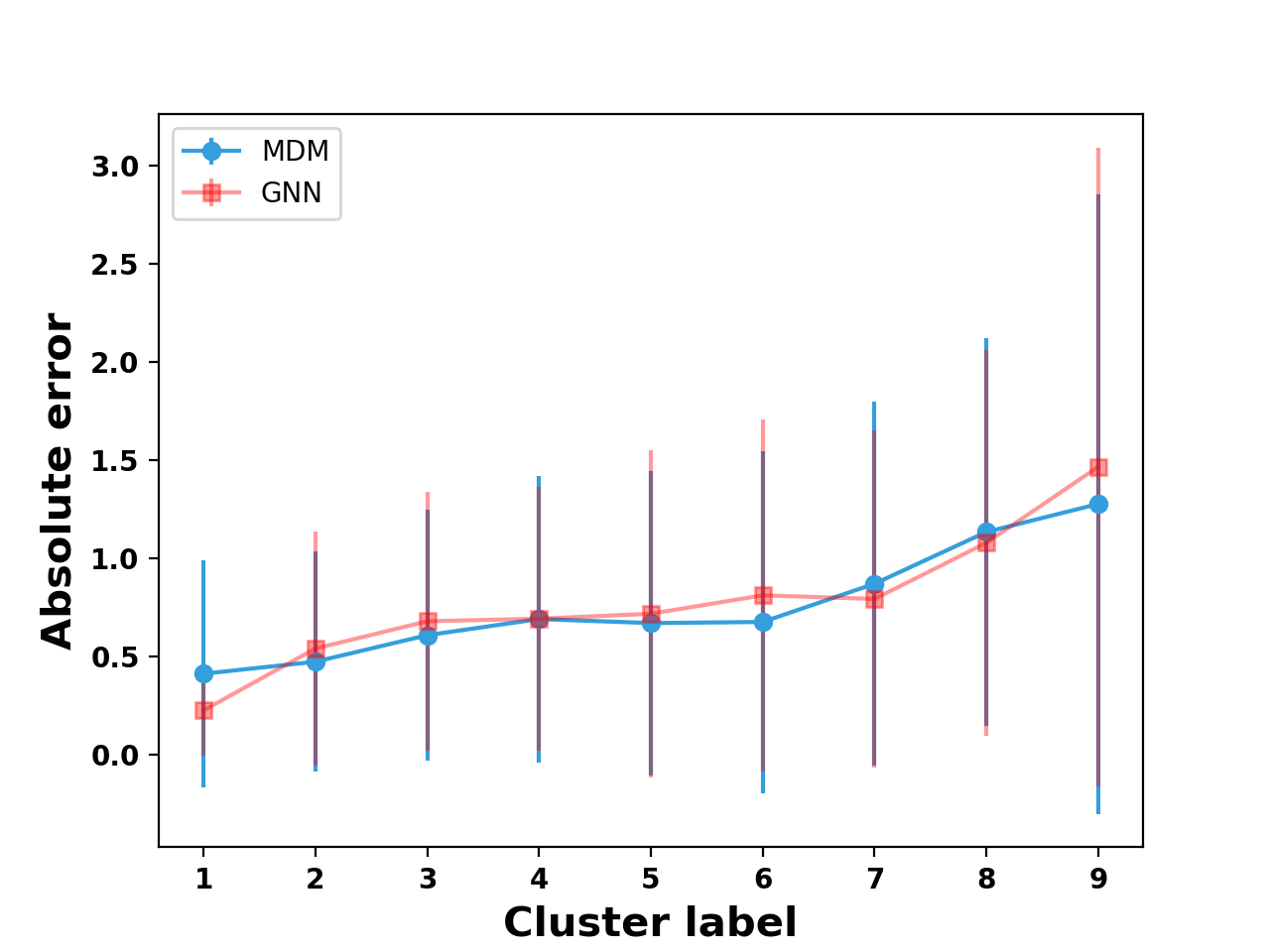}

    \caption{Mean errors by cluster. Error bars indicate the standard deviation across molecules in each cluster.}
    \label{cluster_errors}
\end{figure}%

\subsection{Cluster Analysis}

To better understand what might be driving the patterns in which molecules are easier and harder to predict, we expand our analysis beyond these predefined molecular classes. We would like to analyze whether particular molecular properties influence the predictive ability of the models. We first checked whether the model errors are correlated with any of the molecular features and found that the highest Pearson correlation coefficient was fairly low at around 0.3.

To move beyond analysis at the individual feature level, we aim to determine groups of similar molecules and compare the achieved error levels on these groups. To identify groups of similar molecules, we apply k-means clustering with $k=15$ clusters using molecular descriptors of molecules in the test set. We scaled all the features to zero mean and unit variance to ensure the differing magnitudes of different features does not cause certain features to be more influential in the clustering. We drop six of the resulting clusters that contain less than 10 members. The test errors of the remaining 9 clusters are plotted in Figure \ref{cluster_errors} in ascending order of mean absolute errors of the MDM and GNN models. For each cluster, the ten molecules closest to the cluster center are shown in Figure S12. We note that for the majority of clusters, the two different modeling and representation approaches show very similar error patterns across the groups. This reinforces our earlier conclusion that despite the difference in information available to the two models, they are able to learn similar structure-property relationship patterns. 


We observe that there are significant mean error differences across the different clusters and seek to explain which molecular properties of clusters can best explain observed differences in their error levels by looking for correlations between the average errors across clusters and the average molecular descriptors across clusters. Correlation values of highly correlated features with the error are given in Figure S13. Scatter plots of averaged property values with respect to averaged error are shown in Figure S14. We first observe that the cluster errors do not appear to be driven primarily by molecular size, with a correlation of only 0.48 between average error and number of atoms.  We do find a moderate negative correlation of mean cluster error with mean cluster solubility (-0.65). This observation reinforces results in Figure \ref{err_by_range}, which shows molecules with low solubilities are more difficult to predict.

The descriptors \textit{*C(C)=O} and \textit{cenM9} show the highest correlation with the average cluster errors, with Pearson coefficients of 0.95 and 0.92 respectively. \textit{*C(C)=O} is the count of \textit{*C(C)=O} fragments in the molecule. \textit{cenM9} is a descriptor that quantifies the shape of the molecule and is defined as the $9$\textsuperscript{th} statistical moment of the distribution of distances between the centroid and all the atomic positions of a molecule. Another descriptor that has a high positive correlation with the cluster error is \textit{SRW05}, which is defined as the number of self-returning walk counts of length 5 in the molecular graph. Such self-returning walks can only exist in the presence of 3- or 5-membered rings, with higher values for molecules with a greater number of such rings. The features \textit{cenM9} and \textit{SRW05} can be thought of as measures of the complexity of a molecule. Therefore, it seems that the more complex the molecular structure, the more difficult it is to make predictions for such molecules.




\begin{figure}[t]
    \centering
    \includegraphics[width=.5\textwidth]{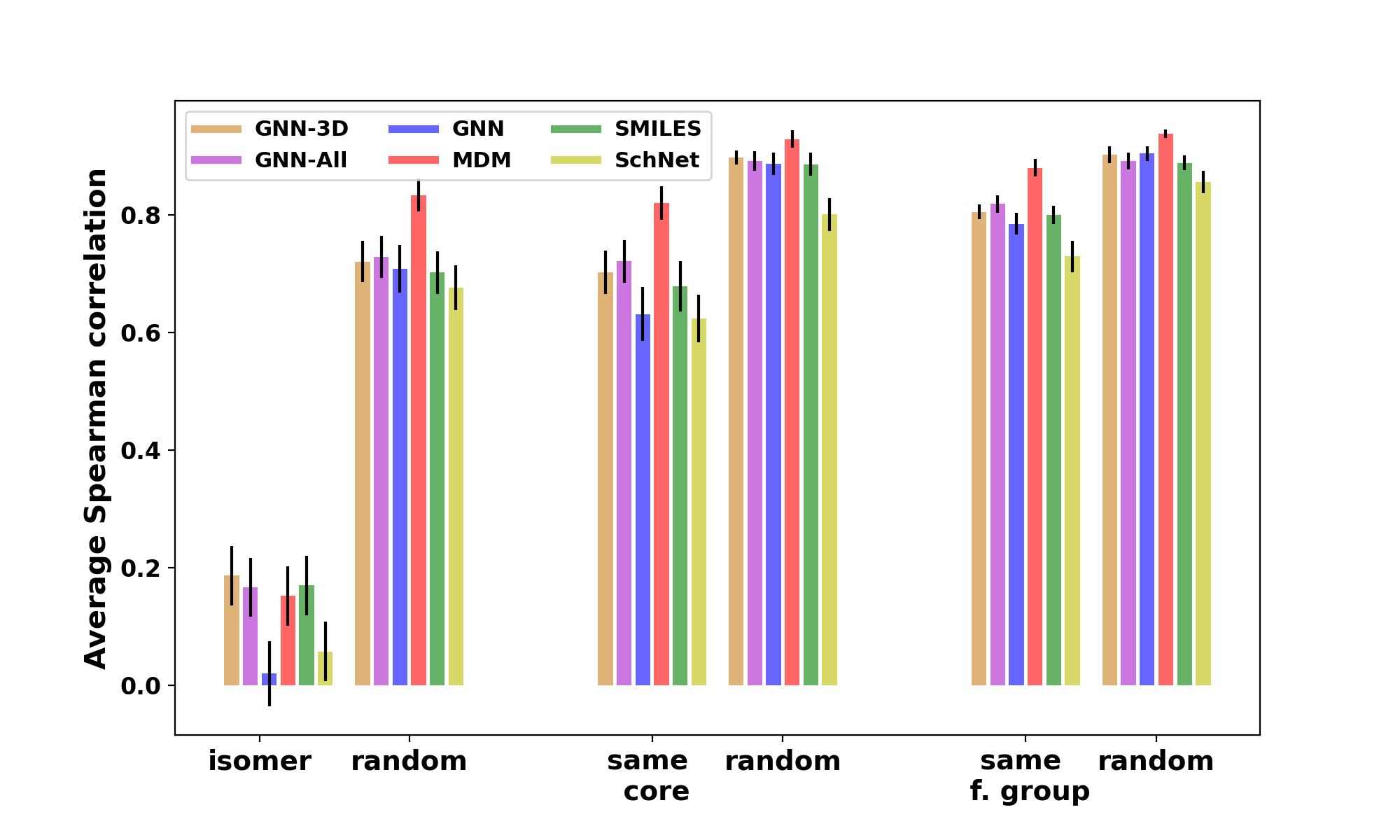}
    \caption{Spearman correlation of actual and predicted solubilities in groups of similar molecules compared with groups of random molecules. We show results for the four main models (GNN, MDM, SMILES, and SchNet) as well as two GNN variants (GNN-3D and GNN-All) discussed in Section~\ref{sec:featureanalysis}.}
    \label{fig:spearman_for_sets}
\end{figure}

\begin{figure*}[t]
\centering
\includegraphics[width=1\textwidth]{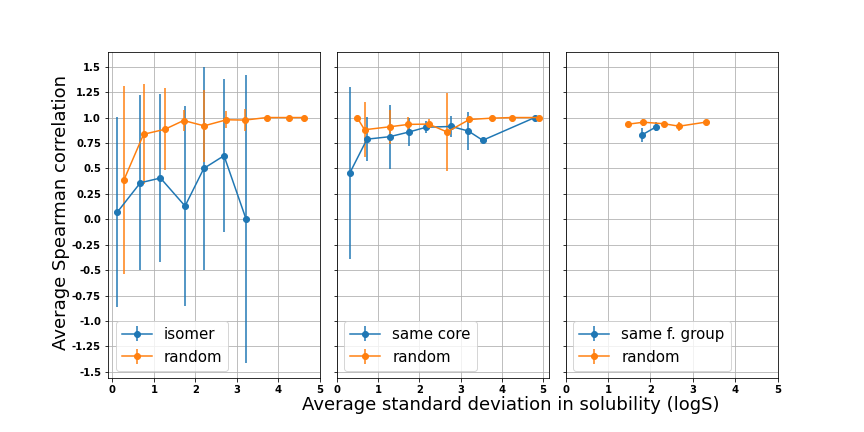}
\caption{Spearman correlation versus the within-group standard deviation for isomer/same core/same functional groups for the MDM model.}
\label{spearman_sdev}
\end{figure*}

\subsection{Molecule Group Evaluation}
\label{sec:groups}

We next analyze the ability of the models to accurately distinguish solubilities of structurally similar molecules. For this analysis, we considered three sets of molecules: (1) positional isomers, (2) molecules with same core structures but different functional groups, and (3) molecules containing same number and type of functional groups attached to different core structures. For example, there are 468 groups of molecules in the isomer set, where each such group consists of $n$ molecules that are isomers of each other. Correspondingly, there are 176 groups of molecules with the same core structure (we excluded isomers from this set) and 21 groups of molecules having the same number and type of functional groups but different core structures.
The median number of molecules in isomer, same-core, and same-functional-group sets are 2, 4, and 37 respectively.  

For each sub-group of similar molecules, we calculated the Spearman correlation coefficient between the predicted and actual solubility values. This measure indicates whether the models are able to correctly rank the molecules within the group from highest to lowest solubility. We then average the Spearman correlation across all sub-groups with each of the three sets. The averaged Spearman correlation for each set is shown in Figure~\ref{fig:spearman_for_sets}. We compare the Spearman correlation observed for these groups of molecules with the correlation achieved for randomly selected groups of molecules of the same size.  We find that, for the same core and functional group sets, the MDM model is able to correctly rank molecules almost as well as it can for random groups of molecules. This is a particular strength of that model over the other three architectures.

However, the ability to rank order the solubilities of molecules in the isomer set is significantly more challenging compared with the other two sets. This result could potentially be explained using the fact that the solubilities in the isomer set do not vary as much as those in a randomly chosen sample (see Figure S8). However, in Figure \ref{spearman_sdev}, we show the Spearman correlation between predicted and actual solubility values versus the level of variability within the group of molecules (as measured by the standard deviation). We see that the Spearman correlation is significantly lower for groups of isomers than for groups of random molecules even after controlling for the level of solubility variation within the group. This shows that the ability to distinguish the effect of functional group positioning on solubility is a key area of improvement for future modeling efforts.

\begin{table*}[t]
\centering
\begin{tabular}{llrrr}
\toprule
\textbf{Model} & \textbf{Features} &   \textbf{R\textsuperscript{2}} &   \textbf{RMSE} &     \textbf{Spearman} \\
\midrule

MDM &DFT\textsuperscript{1} & 0.608 &  1.323 &  0.763 \\
&Mol.\textsuperscript{1} & \textbf{0.750} &  \textbf{1.056} &  \textbf{0.861} \\
&Mol. + DFT\textsuperscript{1} & 0.748 &  1.061 &  0.858 \\
\hline
MDM  &   2D\textsuperscript{2} &  0.766 &  1.066 &  0.876 \\
&   3D\textsuperscript{2} &  0.388 &  1.722 &  0.622 \\
& 2D+3D\textsuperscript{2} &  \textbf{0.772} &  \textbf{1.051} &  \textbf{0.879} \\
\hline
GNN &  w/o 3D coordinates & \textbf{0.74} &  \textbf{1.12} &   \\
    &  with 3D coordinates & 0.69 &  1.21 &  \\
\hline
MetaLayer &  2D & 0.743 &  1.116 &  \textbf{0.862} \\
    &  3D & 0.736 &  1.132 &  0.863 \\
    &  2D+3D & \textbf{0.744} &  \textbf{1.114} &  0.863 \\

\bottomrule
\end{tabular}

\caption{Comparison of model performance when using combinations of features in the MDM and GNN models. \textsuperscript{1}Obtained using the molecules for which both molecular and quantum descriptors are available. \textsuperscript{2}Obtained using the entire dataset. 2D denotes 743 2D descriptors and 59 molecular fragments. 3D denotes 37 3D descriptors and molecular mass. }
\label{tab:dft}
\end{table*}

\subsection{Feature Analysis}
\label{sec:featureanalysis}
We next seek to analyze the importance of different feature types on the ability of the MDM model to accurately predict the solubility. We do this by training alternate versions of the model with certain feature sets added or removed.

While there is a benefit to the development of models that do not depend on inputs requiring computationally and temporally expensive calculations such as density functional theory (DFT), we tested the effect of adding such inputs to our models using a subset of the data for which the quantum descriptors were available. Table~\ref{tab:dft} summarizes the effect of adding these features. It is interesting to note that by using only eight quantum descriptors, the model can achieve reasonable accuracies, even though these accuracies are not as high as those obtained with Mordred-generated molecular descriptors. However, the combination of both quantum mechanical and Mordred-generated features does not result in an improvement compared with the accuracies obtained using the molecular descriptors alone.

We want to understand the importance of 3D molecular shape information on supporting solubility prediction. Therefore, we compare the effect of 2D and 3D descriptors on model performance. In Table \ref{tab:dft}, we list the MDM model accuracies obtained using 2D descriptors alone, 3D descriptors alone, and the combination of both 2D and 3D descriptors. Even with just 2D descriptors, MDM is capable of outperforming our GNN model. The 3D descriptors alone do not have significant predictive power, but do provide a boost in performance when combined with the 2D descriptors. This shows that the 2D and 3D features provide complementary signals related to the structure-solubility relationship.  

The node features of our GNN model depend only on the 2D structural representation of the molecule. As an initial test to check whether incorporating any 3D information have an effect on GNN model accuracy, we added atomic coordinates as node features. These coordinates were generated using Pybel and some molecules were discarded (653, 41 and 57 from train, validation and test sets respectively) after they failed in this generation. We find that adding 3D atomic coordinates as node features does not improve the GNN model performance. Learning the relevant 3D structural features of the compound using atomic coordinates alone as node features seems to be challenging.

An alternate method to add 3D information to the GNN model is to leverage the 3D descriptors as an additional input to the model. Recently, \citet{Peter2018} proposed a graph neural architecture, called the MetaLayer model, which is capable of learning from properties ``global'' to the entire graph structure, which allows us to use the molecular descriptors as an additional input to our GNN model. The results given in Table \ref{tab:dft} shows, consistent with the MDM results, that the 2D descriptors are individually more informative than the 3D descriptors.  However, the addition of the molecular descriptors is not strong enough to surpass the accuracies obtained by our original GNN model.

While the MetaLayer model does not improve on the overall performance, we find that this approach can achieve better accuracies for groups of similar molecules compared to the original GNN model, which may suffer from a lack of 3D information needed to distinguish isomers. These results are shown in comparison with the original GNN model in Figure~\ref{fig:spearman_for_sets}. We see the MetaLayer model that uses only 3D descriptors outperforms all the other models in rank-ordering the solubility values in each isomer group.

\begin{figure*}[!t]
     \centering
     \begin{subfigure}[b]{0.48\textwidth}
         \centering
         \includegraphics[width=\textwidth]{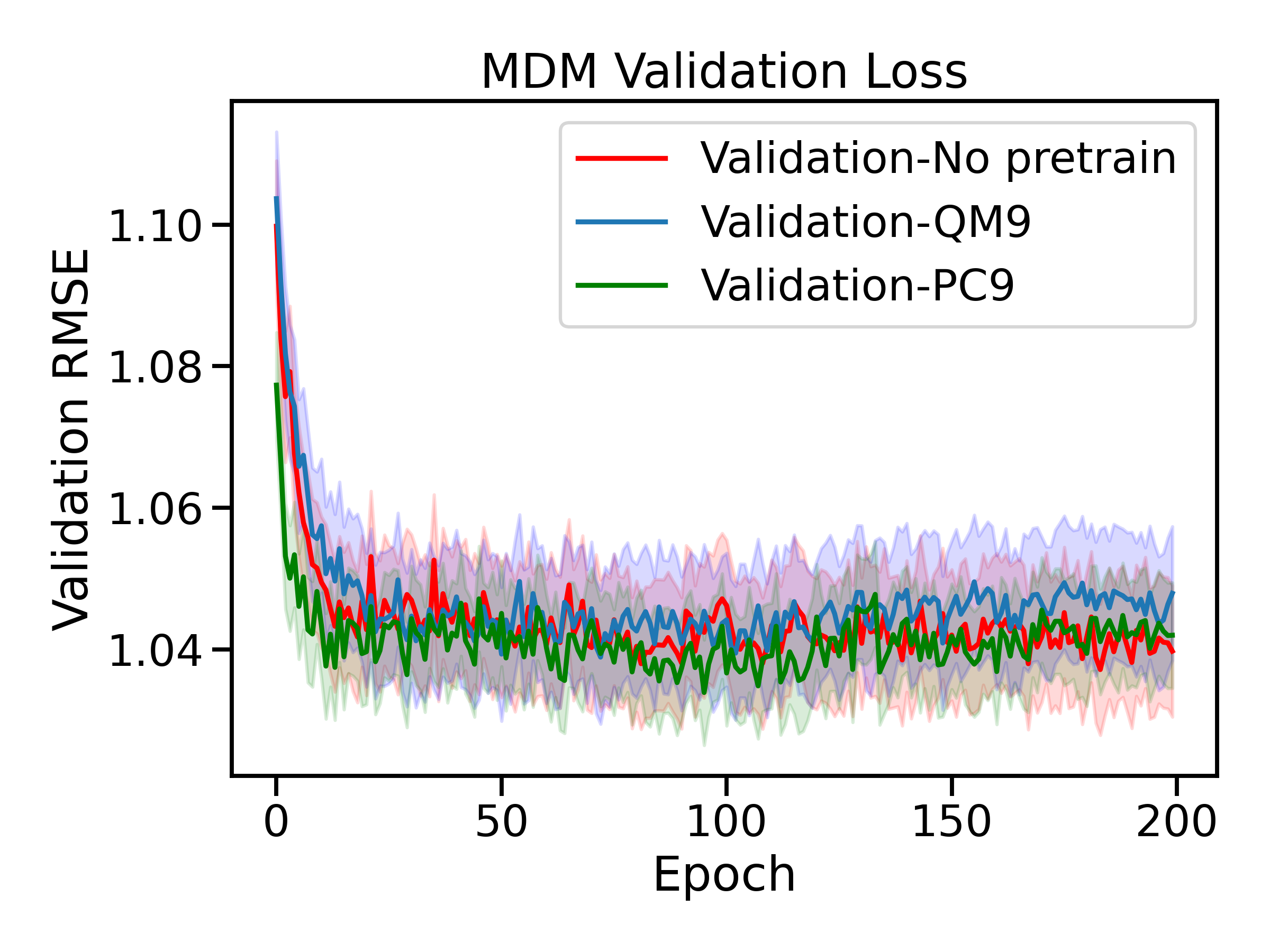}
     \end{subfigure}
     \begin{subfigure}[b]{0.48\textwidth}
         \centering
         \includegraphics[width=\textwidth]{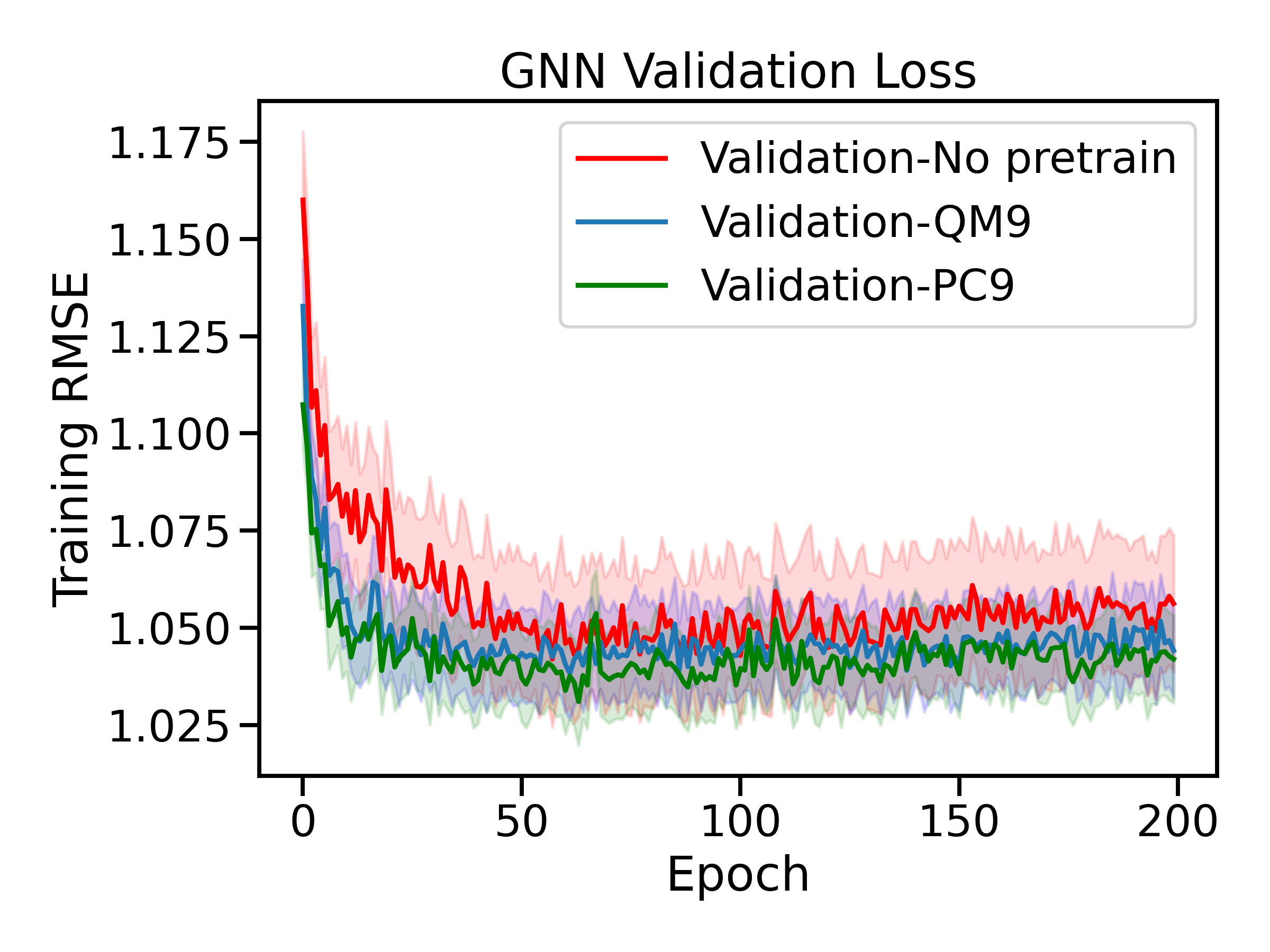}
     \end{subfigure}
        \caption{Learning curves for the MDM model (left) and the GNN model (right) with and without pretraining.}
        \label{fig:pretrain_lc}
\end{figure*}

\begin{figure*}[!t]
     \centering
      \begin{subfigure}[b]{0.48\textwidth}
         \centering
         \includegraphics[width=\textwidth]{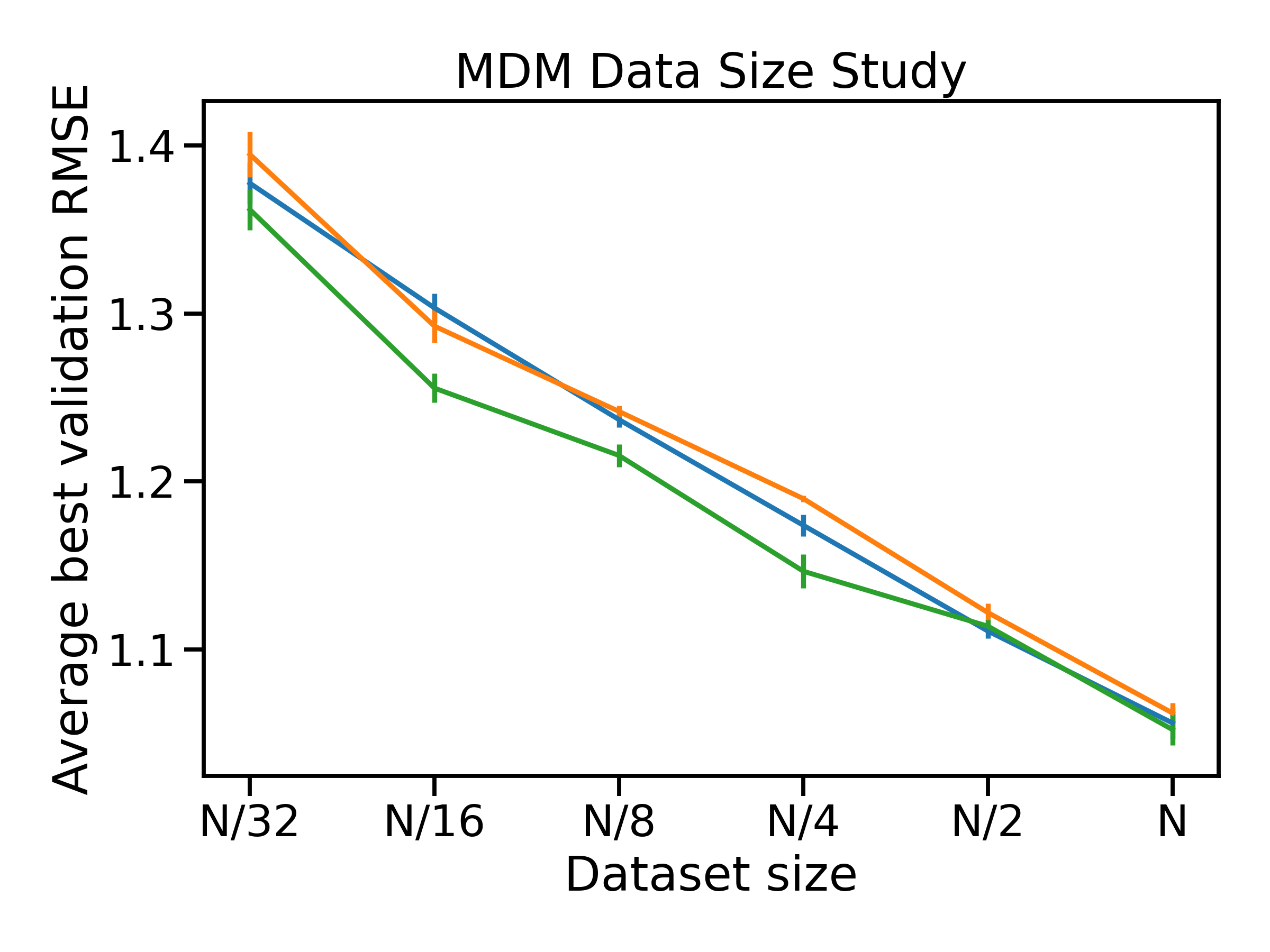}
     \end{subfigure}
     \begin{subfigure}[b]{0.48\textwidth}
         \centering
         \includegraphics[width=\textwidth]{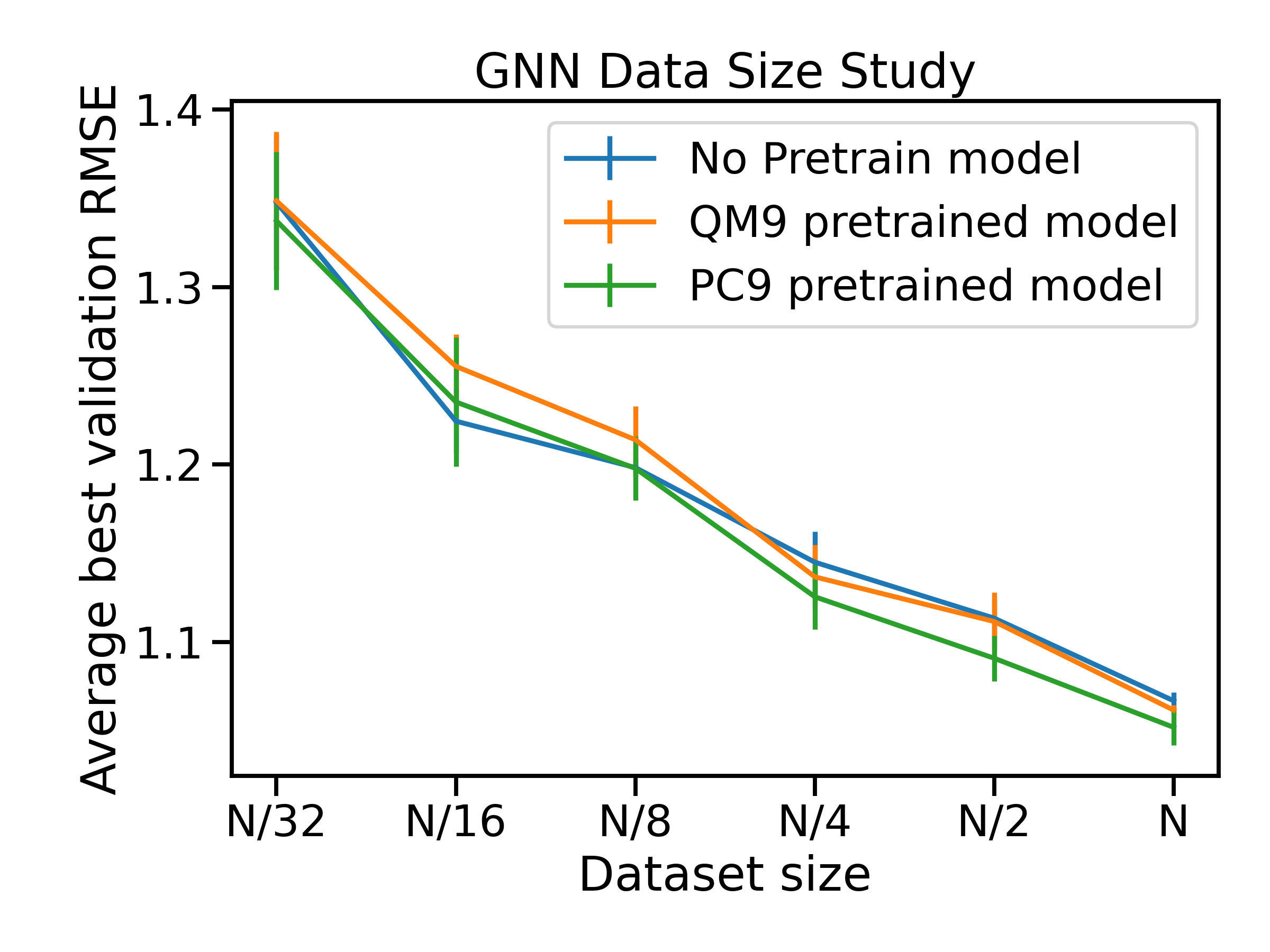}
     \end{subfigure}
        \caption{Model performance (RMSE) as a function of training set size for the MDM model (left) and the GNN model (right).}
        \label{fig:ablation}
\end{figure*}

\section{Effects of Data Size}

While deep learning models have been shown to excel at learning complex patterns like those involved in structure-property relationships, they also typically have large data requirements to achieve good performance at these complex tasks. We perform several analyses to study the impact of dataset size on our model performance. First, we study the impact of transfer learning by pretraining models on large external datasets before fine-tuning on the solubility prediction task. Second, we evaluate our models with smaller subsamples of our data. 

\subsection{Transfer Learning}

Transfer learning is a machine learning technique in which the knowledge a model gains from training on one task is transferred to improve performance on a second task. We apply transfer learning to the solubility prediction task by first pretraining our models on two large datasets, QM9 and PC9.  While these datasets do not contain solubility labels, they are 9 and 7 times larger than our solubility database respectively and can help the model learn patterns that relate molecular structure to molecular properties. To perform transfer learning, we first train MDM and GNN models to predict all the molecular properties included with the QM9 and PC9 datasets and then, starting from weights learned on the QM9 or PC9 dataset, we perform further training using the solubility data.

In Figure~\ref{fig:pretrain_lc}, we show the learning curves of the MDM and GNN models both with and without pretraining, showing how the RMSE decreases during training. We find that pretraining with PC9 data improves the initial performance of the models at the start of training for both the MDM and GNN models.  However, for the MDM model the pretraining on the external data sets does not seem to improve the ultimate achieved performance after fine-tuning.  The GNN model on the other hand, benefits from pretraining with both PC9 and QM9 throughout the training process and pretrained models achieve improved final performance compared with the non-pretrained model. Because the GNN model learns from raw molecular structure while the MDM model learns from pre-derived features, the GNN model benefits more from the additional training data, which can help it learn the complex relationship between raw molecular structure and resulting properties. 

We also observe that across the different results the PC9 dataset provides a bigger boost in performance compared with the QM9 dataset.  This gives evidence for the assertion in~\citet{glavatskikh2019} that the PC9 dataset improves upon the chemical diversity of QM9, leading to better generalization of the patterns learned from the dataset to other datasets and tasks.

\subsection{Data Size Sensitivity}

To investigate the effect of increasing the size of our data set we conducted a data ablation study by decreasing the size of the training data set, with a fixed test set, and analyzed final test accuracy for each training data set size. The data set sizes were calculated by taking the full data set and dividing by increasing integer powers of two, $2^0$, $2^1$, $2^2$, $2^3$, and $2^4$. This results in datasets that are 100\%, 50\%, 25\%, 12.5\%, and 6.25\% of the total size.  We trained the MDM model and GNN model on each data set size in three configurations - from a random weight initialization, from the pretrained QM9 weights, and from the pretrained PC9 weights.  Each model and configuration was trained five times on a given dataset size, using the Adam optimizer with a learning rate = 0.001 for 100 epochs. 

Figure~\ref{fig:ablation} shows the mean and standard deviation of the best validation mean-squared-error for each dataset size both with and without pretraining. We can see the mean-squared-error is still decreasing as the dataset size is increased from half to full, suggesting that increasing our dataset size will continue to improve results. However, it should be noted that as the x-axis is the power of two dividing the full dataset size, this improvement in results will have diminishing returns with respect to the number of data examples added to training. For example, by extrapolating the observed trajectory, we would expect to need to double the training set size to reduce the RMSE below one order of magnitude for the MDM model.

The ablation study also shows some interesting patterns with regard to the combined impact of pretraining and data size. For the MDM, PC9 appears to have a benefit on performance for small solubility datasets but not for large ones. In contrast, the benefit of PC9 pretraining appears for larger datasets using the GNN model.  This difference is likely due to the different requirements of the two models. The MDM model needs to learn a transformation from high-level structural descriptors to the target labels, while the GNN needs to learn a transformation from raw structure information to the target labels.  The GNN may need a larger solubility dataset in order to learn to adapt the patterns that it learned from PC9 to the new solubility target.  Meanwhile, the patterns the MDM must learn are simpler so it can quickly adapt the learning from PC9 with a smaller solubility dataset, and, given a large enough dataset, it can eventually learn the structure-solubility relationship well enough that it cannot be improved by pretraining.

\section{Conclusions and Future Work}

We performed a comparison of different deep learning modeling approaches and molecular representations for the prediction of aqueous solubility using the largest set of solubility measurements to date. Through the use of large, diverse datasets combined with deep learning methods, we demonstrate equal or improved performance of existing solubility prediction datasets. Overall, we found the best performing approach leveraged a set of derived molecular features which comprehensively describe the molecular structure rather than approaches which leverage raw molecular structure information directly. This contrasts with previous studies which have shown the power of deep learning for learning structure-property relationships directly from raw structure~\cite{cui2020,tang2020}. Of the models which did rely on raw structure, graph-based molecular representation showed the strongest performance, almost equalling the MDM model in overall performance but showing reduced ability to distinguish the solubilities of similar groups of positional isomers. 

The superior performance of the MDM model is likely due to its ability to make greater use of 3D molecular shape information than the other models through the use of a set of derived 3D descriptors. However, given that the GNN model is the only one that does not use any 3D information, its achieved performance accuracies are noteworthy. Additionally, even though SchNet was designed to harness the structural information from 3D atomic coordinates, it significantly underperformed the other modeling approaches. We suspect the small training set size compared to the datasets originally used for the SchNet model might have played a role in this result.  Computational requirements also limited the hyper-parameter optimization we were able to perform with this architecture.

There are also considerations other than model accuracy in terms of practical implementation of the different models, including speed and efficiency of computation. In addition to its high accuracy, the MDM is fast to train compared to the other modeling approaches which leverage more complex architectures. However, this model requires the generation of molecular features, and if 3D features are to be included, then atomic coordinates are required. The best form of 3D coordinates are the ones obtained experimentally, but this is not tractable for large datasets. The next best alternative is to optimize geometries using first principles calculations. These calculations are time consuming and obtaining these coordinates for large molecules is not practical. Approximated 3D coordinates can be calculated relatively quickly, however these coordinates are often not reproducible.  This could lead to inconsistent results.

In addition to the evaluation of the overall performance of the models, we performed extensive analysis of the errors observed for different modeling approaches. This error analysis leads to several key findings. Models with differing data representations and architectures make highly correlated errors, showing that they are learning similar structure-property relationships. Model errors are lower for molecules with higher solubility and for solubility ranges with larger amounts of training data and higher for more complex molecules. The models struggle to infer the effect of small structural changes, such as functional group position, on the molecular solubility. Contrary with expectations, 3D information about molecular structure has a limited impact on overall model accuracy but does lead to improved, but still limited, performance on solubility prediction for isomer groups.

Our analyses identify several key directions for improving the predictive performance of solubility prediction models. We determined that pretraining models with large external datasets can provide a performance boost for model architectures which rely on raw structural inputs. While we have initially explored only two such datasets, there is potential for significant improvements using even larger supervised or unsupervised pretraining. We have also confirmed that the number of data points available for training plays a significant role in predictive performance, motivating the collection of additional solubility measurements. However, for some solubility ranges, such as those in the lower or higher ranges, gathering more data can be difficult. Targeting data collection to achieve good coverage in the target solubility ranges of interest for a given application will be key. It is also clear that improvements are needed in the prediction of solubilities of very similar molecules and molecules with multiple fragments, which is likely related to both limitations of the available training data and limitations of current molecular representations and architectures. The collection of focused datasets designed to supervise the improved performance on these molecular types as well as the development of novel representations and model techniques should be targeted for achieving performance improvements.

\section{Data and Software Availability}
Majority of the data used for this work will be made publicly available in a separate publication by \citet{2021Gao} The rest of the data  were extracted from a proprietary repository.

\begin{acknowledgement}

This work was supported by Energy Storage Materials Initiative (ESMI), which is a Laboratory Directed Research and Development Project at Pacific Northwest National Laboratory (PNNL). PNNL is a multiprogram national laboratory operated for the U.S. Department of Energy (DOE) by Battelle Memorial Institute under Contract no. DE- AC05-76RL01830

\end{acknowledgement}


\begin{suppinfo}

Descriptors used for MDM model. Comparison of structural properties of differ- ent datasets. Duplicate removal process. Structure-Solubility Exploration. Graph Neural Network Architecture. Binning solubilities for stratified splitting of the database into train/test and validation folds. Hyper-parameter tuning. Molecular Fragments Analysis. Cluster Analysis

\end{suppinfo}

\clearpage
\bibliography{references}
\end{document}


\title{Supplemental Material}

\section{Descriptors used for MDM model}
\subsection{Mordred}
Here we list the 743 features from the Mordred package which we used for the MDM model.
\begin{table}[htbp]
\tiny
\begin{tabular}{llllllll}
\toprule
          Si &        SddsN &      ATSC5pe &       n9Ring &      SMR\_VSA7 &       ATSC4i &     n3aHRing &  EState\_VSA1 \\
    n8FARing &        ATS0p &     ATSC0are &        Xc-4d &           Mpe &       nHBDon &    PEOE\_VSA1 &       ATSC3Z \\
    n5FARing &        NaaNH &         Mare &       AATS0p &            Sm &        MWC09 &         SddC &      ATS1are \\
 EState\_VSA3 &        ATS7p &        NssPH &  nG12FAHRing &            nH &      ATSC1pe &        NsNH2 &         SsBr \\
    n12HRing &       Xpc-5d &        ATS1p &     n11FRing &        ATSC8p &        ATS2v &       ATSC3i &     n10FRing \\
     ATSC2pe &       ATS6pe &          ABC &        SsSeH &        nFRing &        NssSe &        ATS4Z &    n5FaHRing \\
        NdSe &   nG12FaRing &      n8ARing &   SlogP\_VSA9 &  FilterItLogS &  SlogP\_VSA10 &      n12Ring &     ATSC6are \\
    n8FaRing &      NssssGe &    n4FAHRing &        ATS4p &        ATS1pe &     n5FHRing &      n6aRing &        NsPH2 \\
       piPC1 &           Mv &      nBondsO &     Lipinski &         ZMIC1 &         MIC4 &  EState\_VSA7 &         JGI2 \\
      ATSC8m &        ATS5p &         SsLi &       n4Ring &        Xc-3dv &     Diameter &   PEOE\_VSA10 &      n6ARing \\
      ATS2pe &     VMcGowan &      ATS3are &        ATS8v &         ATS2d &         Sare &       ATSC4v &     SMR\_VSA3 \\
      NsAsH2 &      n11Ring &        MWC06 &        ATS4v &     n11FARing &        Xc-5d &        SsssN &       ATS4dv \\
      NssNH2 &        MWC05 &       ATSC1p &        SaasC &   VSA\_EState6 &      Xch-7dv &      ATSC4dv &       AATS0Z \\
        JGI6 &      ATSC0dv &       Radius &  nBridgehead &       nFARing &       NsssNH &     n3AHRing &        SRW10 \\
    SMR\_VSA2 &         GGI8 &       ATS0pe &      n9HRing &           NdS &         GGI7 &      ATS0are &        ATS3Z \\
       ATS0v &        NsssB &          StN &      ATSC3dv &   VSA\_EState4 &    n12aHRing &        ATS6m &     SMR\_VSA5 \\
        MIC2 &      ATS4are &         MIC5 &        MWC10 &         Xc-6d &    nG12HRing &        WPath &       ATSC4m \\
         NtN &         SssS &     n10ARing &         StsC &         ATS0m &        FCSP3 &    n10FHRing &      ATS6are \\
     SsssGeH &         NsBr &       Xpc-4d &       SsssAs &         Xp-7d &         JGI4 &           nI &       ATS3pe \\
     SssSnH2 &        ATS2m &      n3ARing &           Mm &         GGI10 &        SssNH &       ATS1dv &    n10FaRing \\
   PEOE\_VSA5 &        NdssS &     SMR\_VSA1 &        SRW09 &   EState\_VSA8 &   n10FAHRing &    n9FAHRing &     BalabanJ \\
       SssBe &  EState\_VSA5 &         TIC5 &         StCH &       ATSC2dv &       ATSC0m &      Xch-4dv &      SsssPbH \\
      SssssN &       Xp-3dv &         NssS &        ATS6p &        Xc-4dv &         apol &        ATS6Z &        ABCGG \\
     ATSC3pe &        ZMIC0 &       ATSC5d &         GGI6 &          MPC8 &        NsNH3 &           Sv &       SssCH2 \\
  PEOE\_VSA11 &        ATS4i &        NaaCH &       ATSC5Z &        Xp-1dv &        nAtom &     n5aHRing &          SdS \\
      ATSC4Z &          fMF &     n8AHRing &         SsOH &       ATSC8dv &      NsssSnH &       naRing &         JGI5 \\
      ATSC8d &       NssAsH &      n9aRing &      SssGeH2 &         Xp-5d &       NsSnH3 &       ATSC4d &        SRW05 \\
     NssSiH2 &        NssBH &        NsssN &       Xch-5d &     PEOE\_VSA8 &  VSA\_EState3 &        piPC7 &     n7AHRing \\
    mZagreb2 &     n11aRing &       TMWC10 &        SRW02 &    nG12FARing &        C2SP1 &    n12FARing &       NsSiH3 \\
    n4AHRing &       ATSC2i &          AMW &         GGI5 &        ATSC6v &     SMR\_VSA9 &         NdsN &  VSA\_EState2 \\
      ATS7dv &        nBase &      n9ARing &        ZMIC4 &       SssssSi &        piPC3 &      SssssGe &      nBondsS \\
      AATS0m &      n7aRing &    PEOE\_VSA7 &         NaaN &        NdssSe &         MPC2 &        Xc-3d &       ATSC1i \\
      Xpc-6d &           nF &      SssssPb &           Mi &       Zagreb2 &         CIC1 &        ATS3d &         NsOH \\
     n5aRing &      n4aRing &        NaasC &     n9FHRing &        ATSC2p &        SaaaC &     n6aHRing &      ATS7are \\
  PEOE\_VSA12 &      n5ARing &       NsPbH3 &      n7HRing &     n6FAHRing &       Xch-3d &        MWC03 &       nSpiro \\
       MWC02 &     ATSC1are &   SlogP\_VSA3 &      ATS8are &   nG12FaHRing &      AATSC0Z &      SssssBe &      Xpc-6dv \\
      NssssN &         NddC &        ATS8d &     nBondsKD &    SlogP\_VSA6 &         CIC0 &        ATS7Z &        ATS8Z \\
          MW &   nHeavyAtom &      NssGeH2 &        NssBe &         ZMIC2 &    n5FAHRing &       ATSC6i &      SsssSiH \\
        NsCl &      nBondsA &         TIC0 &       NddssS &        ATSC3d &       ATSC6p &        ATS3m &        C4SP3 \\
      NsssCH &        MWC07 &      SsssssP &        ATS0Z &          WPol &     n4FaRing &       n7Ring &         MPC9 \\
     NssssSn &         NtCH &        SRW04 &   PEOE\_VSA13 &      nFaHRing &     n7aHRing &         GGI4 &      ATSC7dv \\
 VSA\_EState7 &        Xp-4d &      NddssSe &        ATS1v &      AATSC0dv &       n8Ring &   n12FaHRing &         MIC0 \\
     nFaRing &       ATSC7i &    n11FHRing &           SZ &         C2SP3 &        SdsCH &     AATSC0pe &        Xp-1d \\
   n9FaHRing &         CIC4 &        MWC08 &      NssSnH2 &       nFHRing &        ATS7i &      ATSC5dv &       NsssAs \\
     ATSC1dv &          Spe &          IC0 &         TIC1 &          CIC2 &        SsssB &           nP &         NsSH \\
       NssNH &       Xp-2dv &         GGI2 &        NsssP &         MPC10 &      ATSC4pe &        NdssC &       ATS0dv \\
        TIC3 &       n3Ring &       ATSC4p &       n5Ring &       Xpc-4dv &       ATS8pe &       AATS0v &      ATSC7pe \\
        NsLi &         NaaO &       nHBAcc &        SaaCH &        ATSC3p &      AATSC0d &   SlogP\_VSA2 &   SlogP\_VSA7 \\
       SaaNH &        JGI10 &       AATS0i &      n4HRing &         ATS6i &    n10AHRing &       Xch-7d &         nRot \\
      ATSC7m &         JGI3 &      nBondsT &        SaaSe &          JGI9 &      n6FRing &  VSA\_EState8 &         JGI8 \\
      NsGeH3 &        SaasN &     ATSC4are &       SsSnH3 &      SMR\_VSA6 &      n8HRing &        SRW06 &     ATSC8are \\
        NssO &       ATSC3m &        SsssP &       SsAsH2 &      n4FHRing &     n5FaRing &     ATSC2are &        Xp-2d \\
     NsssssP &          IC3 &        piPC2 &    n6FaHRing &          SsSH &        ATS1i &      fragCpx &         SdSe \\
       ATS0i &       piPC10 &         SaaO &       SsSiH3 &      n7FHRing &    n7FaHRing &      Xch-3dv &        C3SP2 \\
          nB &       Xp-5dv &   n11FAHRing &        C1SP3 &         SsCH3 &       ATS5pe &      ECIndex &      n9FRing \\
      SssssB &      n10Ring &        SdCH2 &       SddssS &         NdCH2 &      AATSC0p &          nCl &       ATSC1Z \\
\bottomrule
\end{tabular}
\caption{Mordred descriptors}
\label{table1.1}
\end{table}

\begin{table}[htbp]
\tiny
\begin{tabular}{llllllll}
\toprule
       ATS5v &     n6FHRing &        SsPH2 &  GhoseFilter &        ATSC1m &      ATS5are &        SdssC &    n10aHRing \\
      ATSC6Z &           Mp &       Xp-6dv &        C1SP1 &           SsI &       NdsssP &       ATSC7p &       SsssCH \\
       SRW08 &        ATS5d &      NssssPb &      AATSC0i &            MZ &         bpol &     ATSC5are &       ATSC5v \\
       SRW03 &        SRW07 &     n12ARing &      NsssPbH &        ATSC5i &    nAromBond &    LabuteASA &        piPC6 \\
    n7FARing &       SdssSe &      TpiPC10 &   n10FaHRing &        NssCH2 &         SaaS &      AATSC0m &     SsssssAs \\
        TIC2 &   nG12aHRing &        ATS6d &        ATS2i &         JGT10 &       TSRW10 &      n7FRing &       ATSC2Z \\
      SssssC &        ATS4d &      AATS0dv &           nC &         SsNH2 &          NdO &        Xp-3d &     n11ARing \\
     SssPbH2 &          SsF &  EState\_VSA6 &        ATS8p &      n4aHRing &         MIC3 &          NsI &        C1SP2 \\
       SLogP &        ATS0d &    PEOE\_VSA3 &      ATS2are &         NddsN &       SssNH2 &       ATSC3v &        MWC01 \\
       ATS5i &       ATSC7Z &          IC4 &       SsssNH &          CIC3 &     nBondsKS &        NsSeH &       Xch-4d \\
 EState\_VSA9 &     nFAHRing &     NsssssAs &     n11HRing &         piPC8 &         MPC5 &     ATSC7are &       ATSC8i \\
     SsssdAs &     ATSC3are &    PEOE\_VSA9 &        NsCH3 &        ATSC1d &      TopoPSA &         JGI1 &       ATS2dv \\
     Zagreb1 &      n3HRing &      AATS0pe &     SMR\_VSA8 &      n9FaRing &       ATS8dv &        ATS7m &      SsssSnH \\
      Xc-6dv &  VSA\_EState9 &       ATSC0p &      NsssGeH &         SssSe &       Xp-4dv &        NaasN &       ATS3dv \\
 VSA\_EState5 &       ATSC5m &      n8aRing &      NssPbH2 &         nAcid &     SMR\_VSA4 &      AATSC0v &           nX \\
 VSA\_EState5 &        ATSC5m &     n8aRing &    NssPbH2 &        nAcid &    SMR\_VSA4 &      AATSC0v &          nX \\
     ATSC8pe &       SssssSn &     n6HRing &       NaaS &         NdNH &       ATS3p &       NssssC &   n10FARing \\
       ATS2p &        ATSC8Z &     SssSiH2 &      ATS4m &       ATSC7d &      ATSC7v &     n9AHRing &      SdsssP \\
    n6AHRing &         piPC9 &   nG12FRing &       NtsC &       nBonds &      SssAsH &        C2SP2 &      SsGeH3 \\
     BertzCT &         ATS5m &   n11AHRing &     AATS0d &      ATSC6pe &   n4FaHRing &        NaaSe &     n8FRing \\
     nHetero &         ZMIC3 &        MPC4 &        IC1 &      NssssSi &       ATS8i &       ATSC2d &        GGI1 \\
      ATS4pe &           NsF &        SdsN &        nBr &       Xp-7dv &        CIC5 &  EState\_VSA4 &      ATSC0d \\
       piPC5 &      n7FaRing &        JGI7 &     ATSC6d &       ATS6dv &       SssPH &          SMR &         IC5 \\
        MPC7 &      n9FARing &    n12FRing &    n4ARing &      n7ARing &      Xch-6d &        ATS5Z &      ATSC1v \\
  SlogP\_VSA5 &   SlogP\_VSA11 &      n6Ring &       MPC6 &        nRing &   n12AHRing &        ATS8m &     Xch-5dv \\
     nBondsD &        ATSC2m &        MPC3 &      ATS3i &    nG12ARing &    n9aHRing &      NsssdAs &     n5HRing \\
    n6FARing &          GGI3 &          Sp &      ATS7d &    n7FAHRing &       ATS7v &      ATSC0pe &      nHRing \\
   PEOE\_VSA6 &        Xc-5dv &  nG12FHRing &      ATS6v &       ATSC5p &       ATS2Z &       ATS7pe &    n5AHRing \\
       ATS1m &  EState\_VSA10 &       SdssS &     SsPbH3 &       NssssB &          nO &        NdsCH &  n12FAHRing \\
 EState\_VSA2 &       NssssBe &        SaaN &  PEOE\_VSA4 &    AATSC0are &   n11aHRing &          IC2 &    n12aRing \\
        SsCl &     n12FaRing &        SssO &       SdNH &        NaaaC &   n11FaRing &    n8FaHRing &        MIC1 \\
     Xpc-5dv &        ATSC0v &    n10HRing &     ATSC8v &     n10aRing &       C3SP3 &    nAromAtom &        GGI9 \\
       MWC04 &       n3aRing &       ZMIC5 &    ATSC6dv &     nG12Ring &       Xp-6d &       nARing &    n8FHRing \\
          nN &          TIC4 &   PEOE\_VSA2 &    NsssSiH &        ATS1d &  SlogP\_VSA4 &           nS &     naHRing \\
      ATSC6m &    nG12AHRing &      TMPC10 &   AATS0are &        ATS3v &         SdO &   SlogP\_VSA8 &       SsNH3 \\
       piPC4 &     nG12aRing &      ATSC2v &    nAHRing &      Xch-6dv &    n4FARing &   SlogP\_VSA1 &   n8FAHRing \\
 VSA\_EState1 &      n6FaRing &     SddssSe &     ATSC0i &   n11FaHRing &   n12FHRing &      nBondsM &      ATS5dv \\
       SssBH &       n5FRing &    n8aHRing &    n4FRing &  TopoPSA(NO) &       ATS1Z &       ATSC0Z &             \\
\bottomrule
\end{tabular}
\caption{Mordred descriptors continued.}
\label{table1.2}
\end{table}

\subsection{Fragments descriptors}

\begin{figure*}[htbp]
\centering
\includegraphics[width=1.2\textwidth]{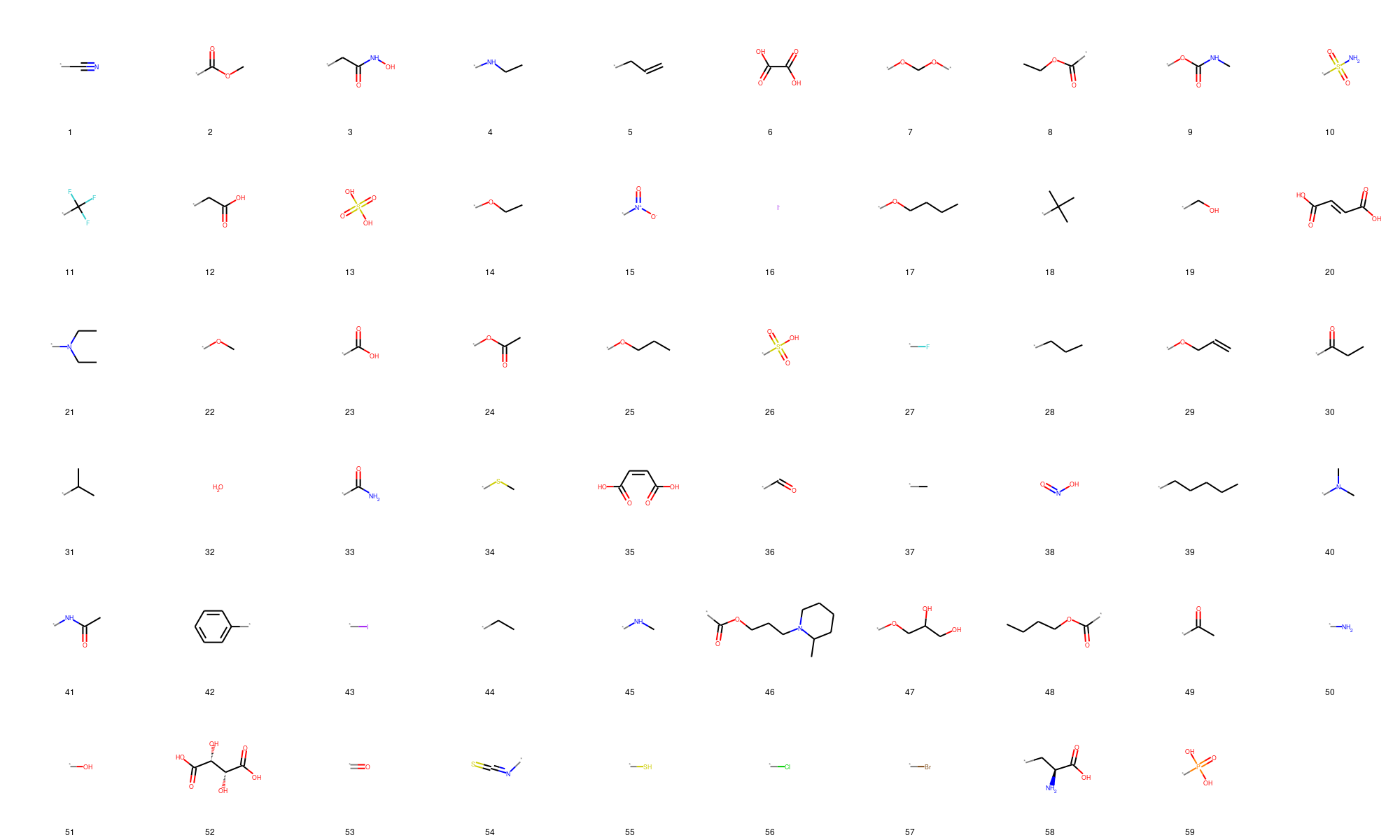}
\caption{Most common fragments determined using RDKit (1-52). Other commonly found fragments identified with the help from a chemistry expert (53-59).  }
\label{}
\end{figure*}

\clearpage

\section{Comparison of structural properties of different datasets}
We present further comparison of the properties of our dataset with external datasets used for performance comparison.  

\begin{table}[!htb]
\begin{tabular}{llllllll}
\toprule
 \textbf{dataset} &       \textbf{Mass} & \textbf{H.Atom} & \textbf{A.Bond} &     \textbf{*Cl} &  \textbf{*C} & \textbf{*=O} &      \textbf{*O} \\
\midrule
    Ours &  16 - 1819 &   1 - 132 &   0 - 66 &  0 - 12 &  0 - 24 &  0 - 17 &  0 - 33 \\
\midrule
 Delaney &   16 - 780 &    1 - 55 &   0 - 30 &  0 - 12 &   0 - 7 &   0 - 6 &  0 - 11 \\
   SAMPN &   46 - 665 &    2 - 47 &   0 - 27 &  0 - 10 &   0 - 7 &   0 - 6 &   0 - 8 \\
     Cui &  16 - 1583 &   1 - 109 &   0 - 64 &  0 - 12 &  0 - 24 &  0 - 17 &  0 - 20 \\
\bottomrule
\end{tabular}
\caption{Additional structural properties of molecules in the datasets used in this work. H.Atom and A.Bond are the counts of heavy atoms atoms and aromatic bonds. *Cl,*C,*=O, and *O are the counts of fragments containing -Cl, -C, =O and -OH, where "*" denotes any arbitrary atom.} 
\label{}
\end{table}

\begin{figure*}[!htb]
\centering
\includegraphics[width=1.0\textwidth]{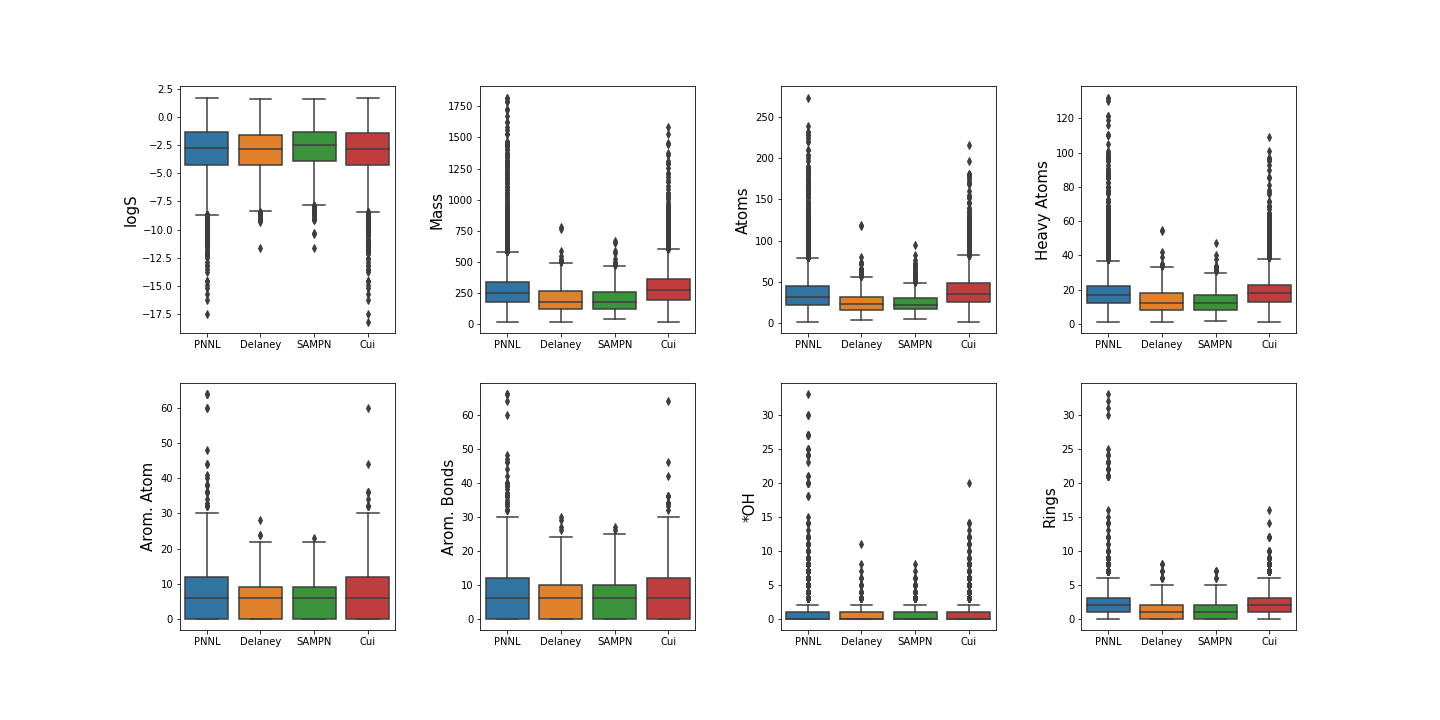}
\caption{Box plots showing the distribution of different structural properties of PNNL, Delaney, Tang, and Cui datasets.}
\label{}
\end{figure*}
\clearpage

\section{Duplicate removal process}
Because we perform analysis of the model performance on different combinations of our data with external datasets, we must deal with duplicate entries that exist across the datasets. We used a process similar to what is described in \cite{sorkun_2019} to resolve duplicates in the external datasets. If the number of duplicates is exactly two and the difference between these solubilities is less than 0.03, two entries were merged by considering the average of the two values. If the difference is greater than 0.03, the values were discarded. If the number of duplicates are greater than two, their standard deviation was calculated. If the standard deviation is less than 0.05, the solubility closest to the mean solubility of the duplicates was kept in the dataset and the other were discarded. Next, the SMILES were converted to the format defined in RDKit in order to make sure that all the SMILES strings across all the datasets conform to the same convention. If any duplicate SMILES resulted after this conversion, all such duplicates were discarded as the duplication might have been caused by limitations in RDKit. This process discarded 22, 2 and 8 molecules in Delaney, Huuskonen and Solubility Challenge 1 datasets respectively. In the Delaney dataset, additional 11 pairs of SMILES strings which were detected to be duplicates after being read by RDKit, were also removed. In the Huuskonen dataset, 4 SMILES strings failed to be read by RDKit and another 2 pairs of SMILES happened to be duplicates after converting to the canonical form.

\section{Structure-Solubility Exploration}
\label{sec:structure_property}

\begin{figure}[t]
    \centering
    \includegraphics[width=.9\textwidth]{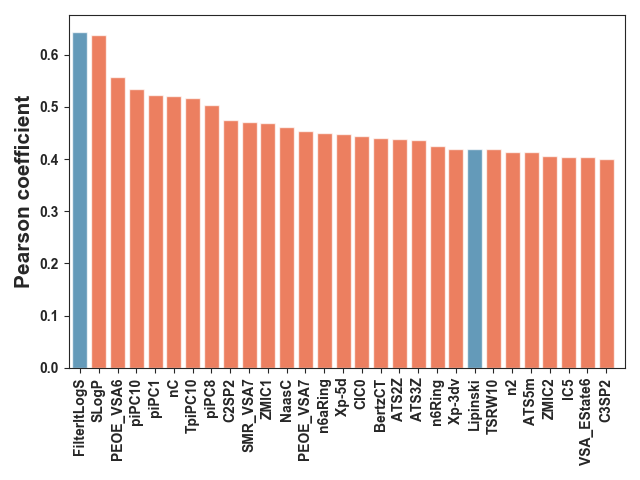}
    
    \caption{Absolute value of Pearson correlations of highly correlated features with log solubility with positive correlations in blue and negative correlations in red}
    \label{fig:sol_corr}
\end{figure}%

\begin{figure}[t]
    \centering
    \includegraphics[width=.9\textwidth]{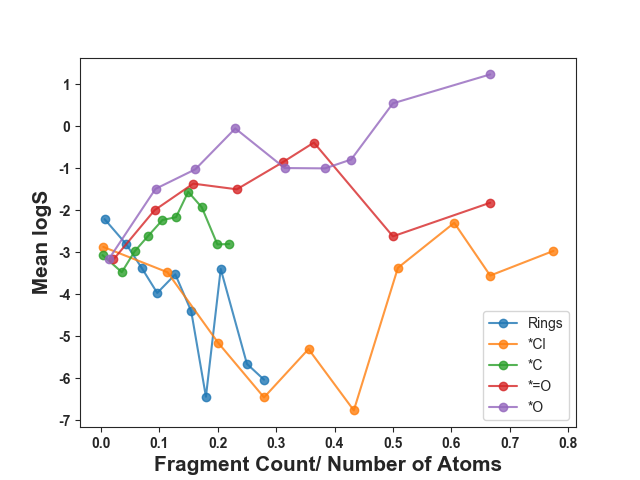}
    \caption{Mean solubility of molecules within different bins of the number of given fragments and rings in the molecules normalized by molecular size.}
    \label{fig:rings}
\end{figure}%

\begin{figure*}[!htb]
\centering
\includegraphics[width=1.1\textwidth]{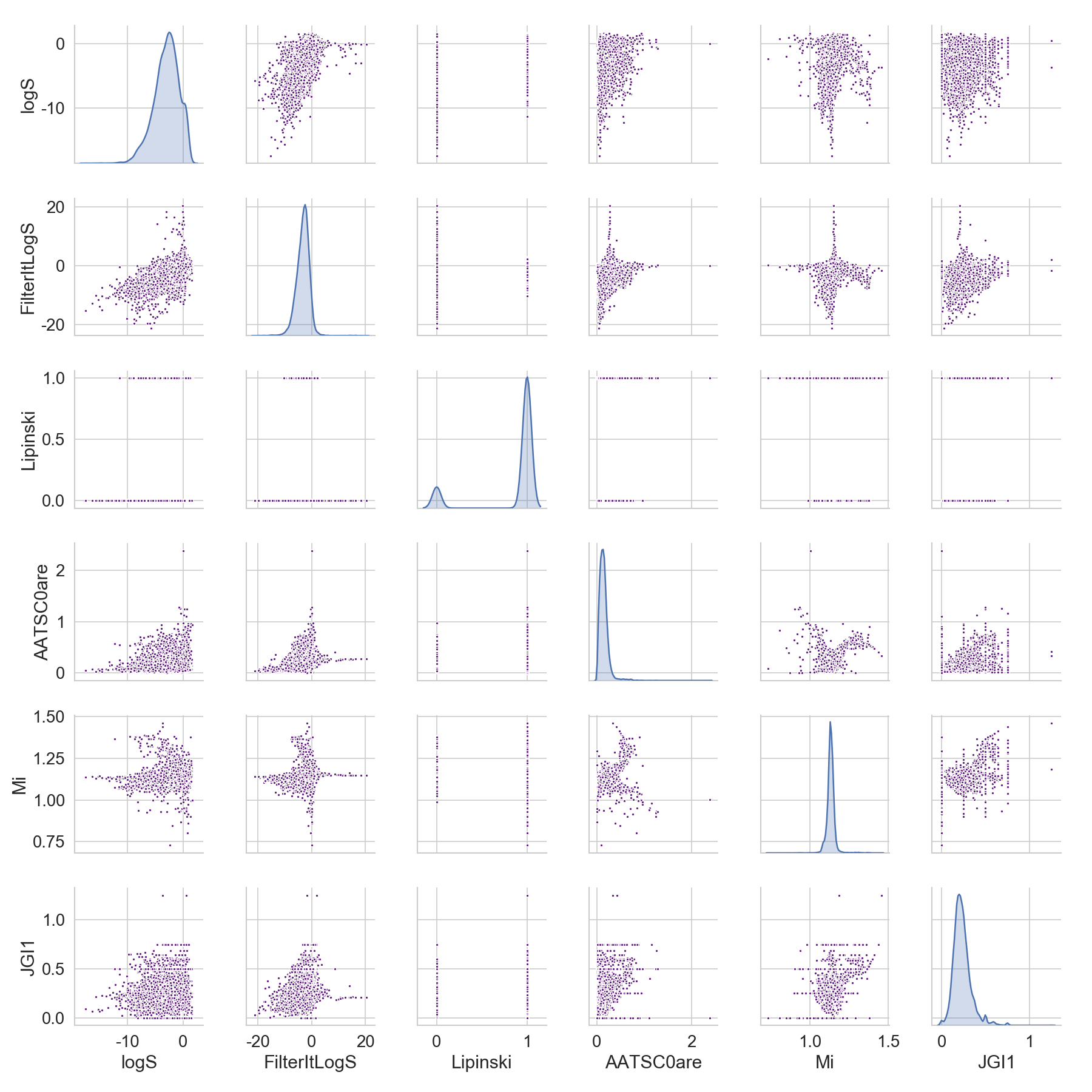}
\caption{Top 5 positively correlated features.}
\label{}
\end{figure*}

\begin{figure*}[!htb]
\centering
\includegraphics[width=1.1\textwidth]{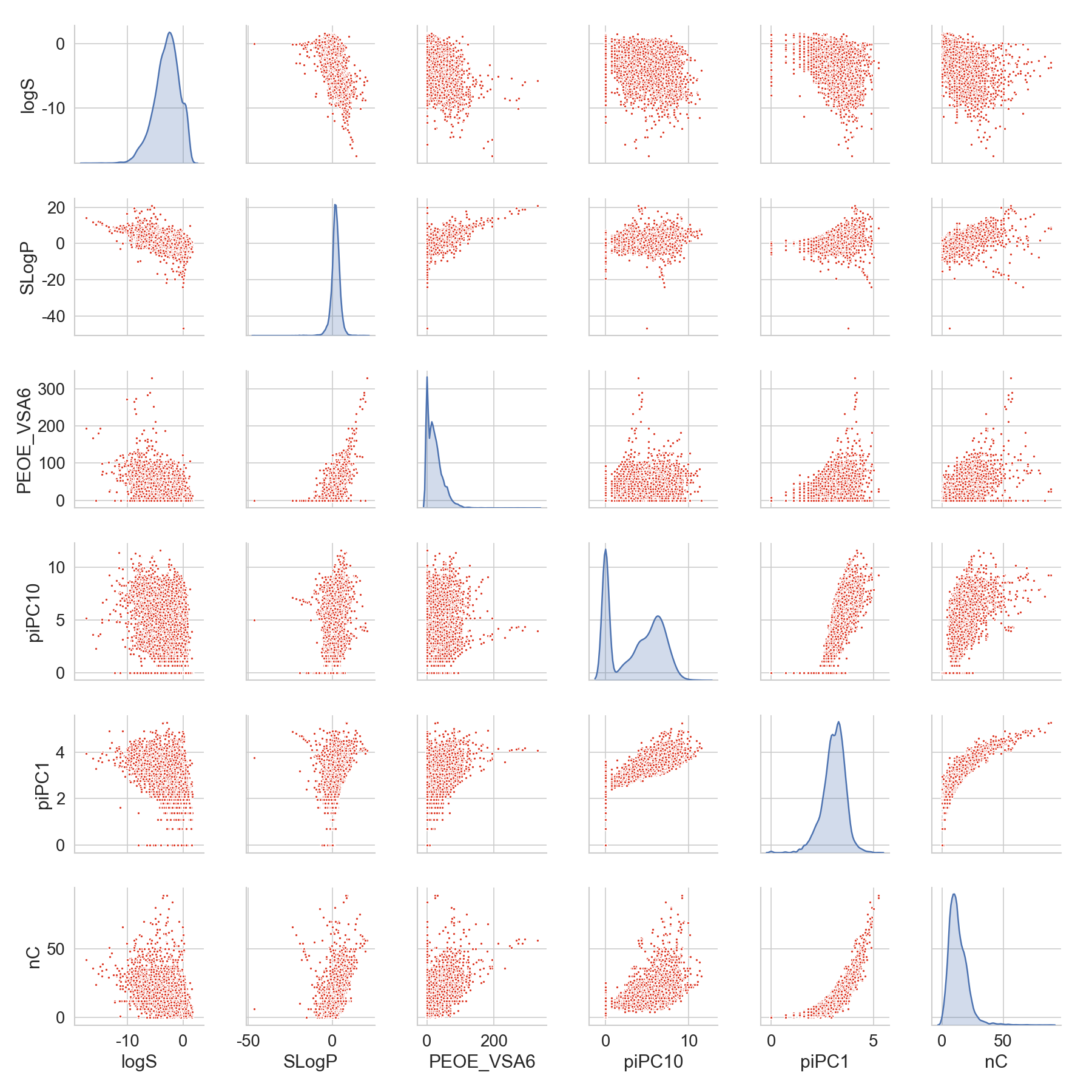}
\caption{Top 5 negatively correlated features.}
\label{}
\end{figure*}

Before applying predictive modeling to our dataset, we first explore the data and perform a structure-property relationship study. We start by analyzing the relationships between measured solubility values and molecular descriptors as calculated by Mordred.
Features that correlate with log solubility with an absolute Pearson correlation coefficient greater than 0.4 are shown in Figure~\ref{fig:sol_corr}. Positive and negative correlations are indicated using blue and red colors respectively.
This result only includes features that do not contain any missing values. Additionally, because we expect many of the molecular descriptors to be highly correlated, we also removed any features with a correlation coefficient greater than 0.95 with another feature while keeping the feature that has higher correlation with solubility. This allows us to identify features which likely provide independent predictive signals of solubility. \textcolor{dgreen}{Pairwise scatter plots of most positively and negatively correlated descriptors are shown in Figure S5 and Figure S6.} We next describe some of the 29 molecular descriptors identified to have an absolute correlation greater than 0.4.

\textit{FilterItLogS} is a theoretical approximation for solubility originally used in the Filter-it software, which explains why this feature has such high observed correlation with solubility. \textit{FilterItLogS} is calculated as, 

\begin{equation}
\mathrm{FilterItLogS} = 0.898 + 0.104 \sqrt{\mathrm{MOLWT}} + w_i c_i,
\end{equation}

where MOLWT is the molecular weight, and $w_{i}$ is the weight corresponding to the count of $i^{th}$ fragment $c_{i}$. The types of fragments used are listed in Table \ref{filtert_logS_fractions} along with the corresponding weights used in the RDKit implementation of \textit{FilterItLogS}.

\begin{table}[!htb]
\centering
\begin{tabular}{ll}
\toprule
 \textbf{Fragment} &       \textbf{Weight} \\
\midrule
    $[NH0;X3;v3]$ & 0.71535 \\
    $[NH2;X3;v3]$ & 0.41056 \\
    $[nH0;X3]$ & 0.82535 \\
    $[OH0;X2;v2]$ & 0.31464 \\
    $[OH0;X1;v2]$ & 0.14787 \\
    $[OH1;X2;v2]$ & 0.62998 \\
    $[CH2;!R]$ & -0.35634 \\
    $[CH3;!R]$ & -0.33888 \\
    $[CH0;R]$ & -0.21912 \\
    $[CH2;R]$ & -0.23057 \\
    $[ch0]$ &  -0.37570 \\
    $[ch1]$ & -0.22435 \\
    $F$ & -0.21728 \\
    $Cl$ & -0.49721 \\
    $Br$ & -0.57982 \\
    $I$ & -0.51547 \\
\bottomrule
\end{tabular}
\caption{Fragments and weights used for FilterItLogS}
\label{filtert_logS_fractions}
\end{table}

In general, the fragments containing N and O have positive weights and those containing C and halogen atoms have negative weights. Therefore, the high correlation of \textit{FilterItLogS} with solubility implies that the solubility is proportional to the counts of O and N containing fragments and inversely proportional to the molecular weight and the number of halogens in the molecule.

\textit{SlogP} is the octanol-water partition coefficient  calculated based on the method proposed by Wildman and Crippen~\cite{Wildman1999}. This correlation is also not surprising as the octanol-water partition coefficient has been known to correlate with solubility~\cite{Hansch1968,delaney2004}. \textit{PEOE\_VSA6}, \textit{SMR\_VSA7} and \textit{PEOE\_VSA7} are measures of van der Waals surface area of the atoms. The fact that these terms having negative correlations with solubility indicates that the larger the size of the molecule, the less soluble it is. In fact, the molecular size is considered as an important feature for solubility prediction~\cite{Hewitt2009}. 
\textit{piPC1}, \textit{piPC8}, and \textit{piPC10} are path counts (that are weighted by bond order)  of length 1, 8 and 10 respectively. \textit{TpiPC10} is sum of weighted path counts over the path lengths 1 to 10. \textit{nC} is the number of Carbon atoms. \textit{C2SP2} is the number of SP2 carbon atoms bound to two other carbons. \textit{ZMIC1} is a measure of the information content (Shannon's entropy) calculated by classifying atoms based on the bond order and the type of the neighboring atom. \textit{NaasC} is the number of carbon atoms to which two aromatic bonds and a single bond are attached. \textit{n6aRing} is the number of 6-membered aromatic rings in the molecule. \textit{Xp-5d} is the Chi connectivity index (weighted by sigma electrons) for fragments containing 5 bonds. \textit{CIC0} is the complementary information content based on different types of atoms in the molecule.
\textit{BertzCT} is a measure of ``complexity" of a molecule. According to RDKit's documentation, this feature consists of two parts to quantify the complexity of bonding and the distribution of heteroatoms.

In the Mordred implementation, the \textit{Lipinski} value for a given molecule is a binary indicator of whether all four of Lipinski rules are satisfied (number of Hydrogen bond donors $\leq$ 5, number of Hydrogen bond acceptors $\leq$ 10, molecular weight $\leq$ 500, and logP $\leq$ 5). According to Lipinski, an orally active drug should satisfy at least three of these rules. Therefore, some information regarding solubility should be embedded in these rules. However, as seen from Figure S3 many molecules with high measured solubility have Lipinski values of zero.

Another way to explore the trends in solubility is in terms of molecular fragments. Early work on solubility often used the counts and fractions of fragments as a predictive signal. As we saw with the \textit{FilterItLogS} descriptor, such features have high correlation with the molecular solubility.

\begin{figure*}[h]
\centering
\includegraphics[width=1.1\textwidth]{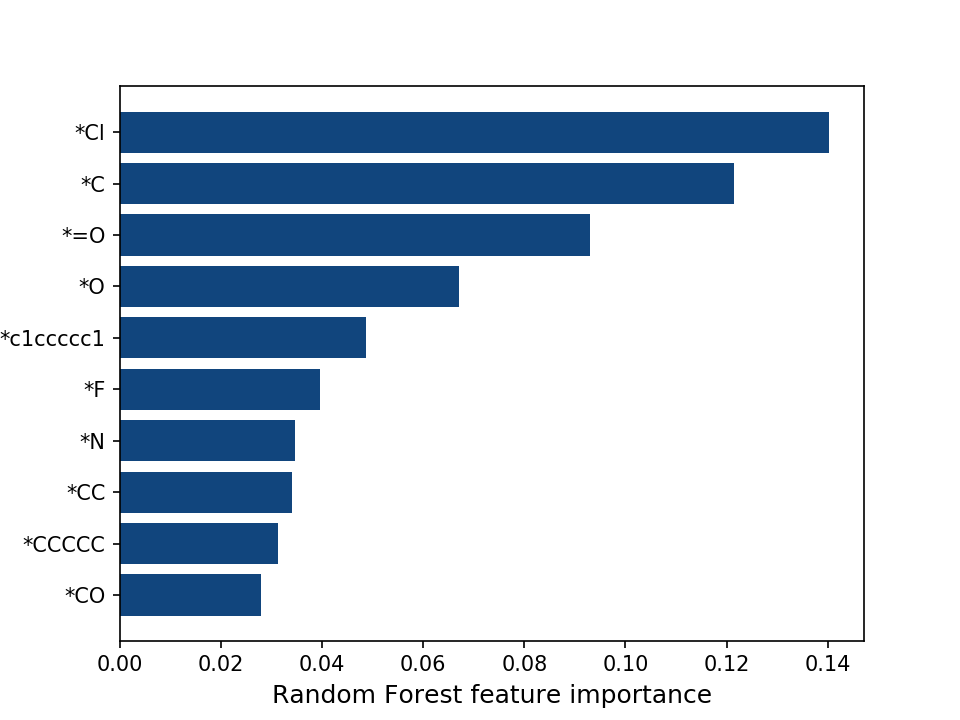}
\caption{Feature importance of top 10 highly important fragments as determined by the random forest algorithm. }
\label{fig:rf-fimp}
\end{figure*}

 We trained a random forest model using the log solubility as the target property and the counts of 59 fragments (described in the Data section of the main text) within the molecule as the features. \*Cl, \*C, \*=O, and \*O are among the highly influential features according to random forest's feature importance metric (the feature importance scores of the most important ten features are given in Figure \ref{fig:rf-fimp}). In Figure \ref{fig:rings}, we show how the solubility changes with respect the prevalence of these four fragments within a molecule. Since the raw fragment counts are likely to be correlated with the size of the molecule (which also affects the solubility), we normalize the molecular fragment counts by the total size of molecule in terms of number of atoms. Not surprisingly, O containing fragments have positive impact on the solubility; the higher the number O-containing fragments the higher the solubility. The O containing fragments are instrumental in forming hydrogen bonds with water during the solvation process. \*C denotes C-A single bonds, where A can be any atom. The presence of single bonds involving C appears to be favourable for solubility, when at most the count of such bonds is less than 20\% of the number of atoms in the molecule. Cl containing fragments seem to have mixed effects on solubility. As long as the number of Cl atoms constitute less than 40\% of the total number of atoms (assuming that Cl-Cl fragments do not exist), an increase in the Cl content of the molecules results in a reduction in solubility. This is expected as halogen bonds are known to be hydrophobic~\cite{Priimagi2013}. However, when the Cl content increases further we see the the solubility of the corresponding molecules increase. This is probably due to the effect of other atoms and functional groups in the molecule.
 
As 80\% of the molecules in our dataset contain rings, it is also interesting to analyze how solubility is related to the number of rings. In addition to the specified fragments discussed above, Figure~\ref{fig:rings} also shows the relationship between solubility and the number of rings relative to the molecules size. We find that molecules that have higher numbers of rings relative to their size tend to have reduced solubility. 




\begin{figure}[t]
    \centering
    \includegraphics[width=.9\textwidth]{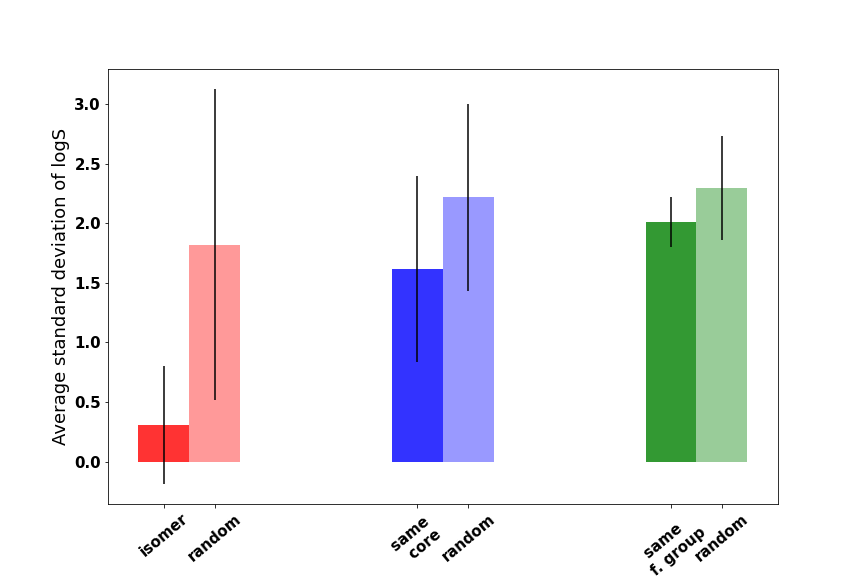}
    \caption{Average standard deviation of solubilities of molecules in sets of isomer, same core, and same functional groups.}
    \label{sdev_vs_groups}
\end{figure}%

In addition to the presence of different molecular fragments, the position and structure of fragments/functional groups can also have a significant effect on the solubility. As we will show later in the text, predicting the solubility of positional isomers is a challenging task. In order to understand the level of solubility variation among groups of similar molecules, we considered three sets of molecules: (1) positional isomers, (2) molecules with same core structures but different functional groups, and (3) molecules containing same number and type of functional groups attached to different core structures. For example, there are 468 groups of molecules in the isomer set, where each such group consists of $n$ molecules that are isomers of each other. Correspondingly, there are 176 groups of molecules with the same core structure (we excluded isomers from this set) and 21 groups of molecules having the same number and type of functional groups but different core structures.
The median number of molecules in isomer, same-core and same-functional-group sets are 2, 4 and 37 respectively. In Figure~\ref{sdev_vs_groups} we compare the level of solubility variability that exists in these groups relative the level of variability in random groups of molecules of the same size.  We see that in all cases the groups of similar molecules had less variation in solubility than random groups of molecules. However, the level of variability among the isomer groups was significantly lower than the other two group types.

\section{Graph Neural Network Architecture}

The graph neural network architecture leveraged in this work uses an iterative process called message passing to update the node and edges features during training. At each iteration, node features of node $i$ ($\mathbf{x}_{i}$) are updated according to, $\mathbf{x}_{i}^{t} = \gamma^{t-1}(\mathbf{x}_{i}^{t-1}, m_{i}^{t-1})$, where  $\gamma^{t-1}$ is the update function which is a differentiable function like a multi layer perception and $m_{i}^{t-1}$ is the aggregated message coming from the neighboring nodes given by

\begin{equation}
m_{i}^{t-1} = \Lambda_{j \in \mathcal{N}(i)} \, \phi_{\mathbf{\Theta}} \left(\mathbf{x}_i, \mathbf{x}_j,\mathbf{e}_{j,i}\right).
\end{equation}

$\mathcal{N}(i)$ are the neighboring atoms to atom $i$.
$\Lambda$ is a differentiable function that is used to aggregate the message of a given node with those of its neighboring ones. Usually, this function is one of summation, mean, or max. $\phi_{\mathbf{\Theta}}$ is another differentiable function like a multi layer perceptron.

For graph convolution networks (GCNs), $m_{i}^{t-1}$ takes the form,
$ \sum_{j  \in N(i)} \frac{1}{\sqrt{\mathrm{deg}(i)}} \frac{1}{\sqrt{\mathrm{deg}(j)}} \Theta \cdot x_{i}^{t-1}$,
and the update function is summation~\cite{KipfW16, Fey19}. Thus the node features in a GCN are updated as,

\begin{equation}
x_{i}^{t} = \sum_{j  \in N(i)} \frac{1}{\sqrt{\mathrm{deg}(i)}} \frac{1}{\sqrt{\mathrm{deg}(j)}} (\Theta \cdot x_{j}^{t-1}), 
\end{equation}
where $\Theta$ is a weight matrix used to linearly transform the node features and $\mathrm{deg}(i)$ is the degree of the $i^{th}$ node.

We also use an edge convolutional layer~\cite{Wang18} for which the edge representations are updated according to,
\begin{equation}
   \mathbf{x}^{t}_i = \sum_{j \in \mathcal{N}(i)}
        h_{\mathbf{\Theta}}(\mathbf{x}_i^{t-1} \, \Vert \,
        \mathbf{x}_j^{t-1} - \mathbf{x}_i^{t-1}).
\end{equation}

Here $h_{\mathbf{\Theta}}$ represents an arbitrary neural network and $||$ denotes concatenation of two vectors.
An illustrated description on the general mechanism of message passing and aggregation in a graph neural network is shown in Figure S6. 


\begin{figure}[tbp]
\centering
\includegraphics[width=1.0\textwidth]{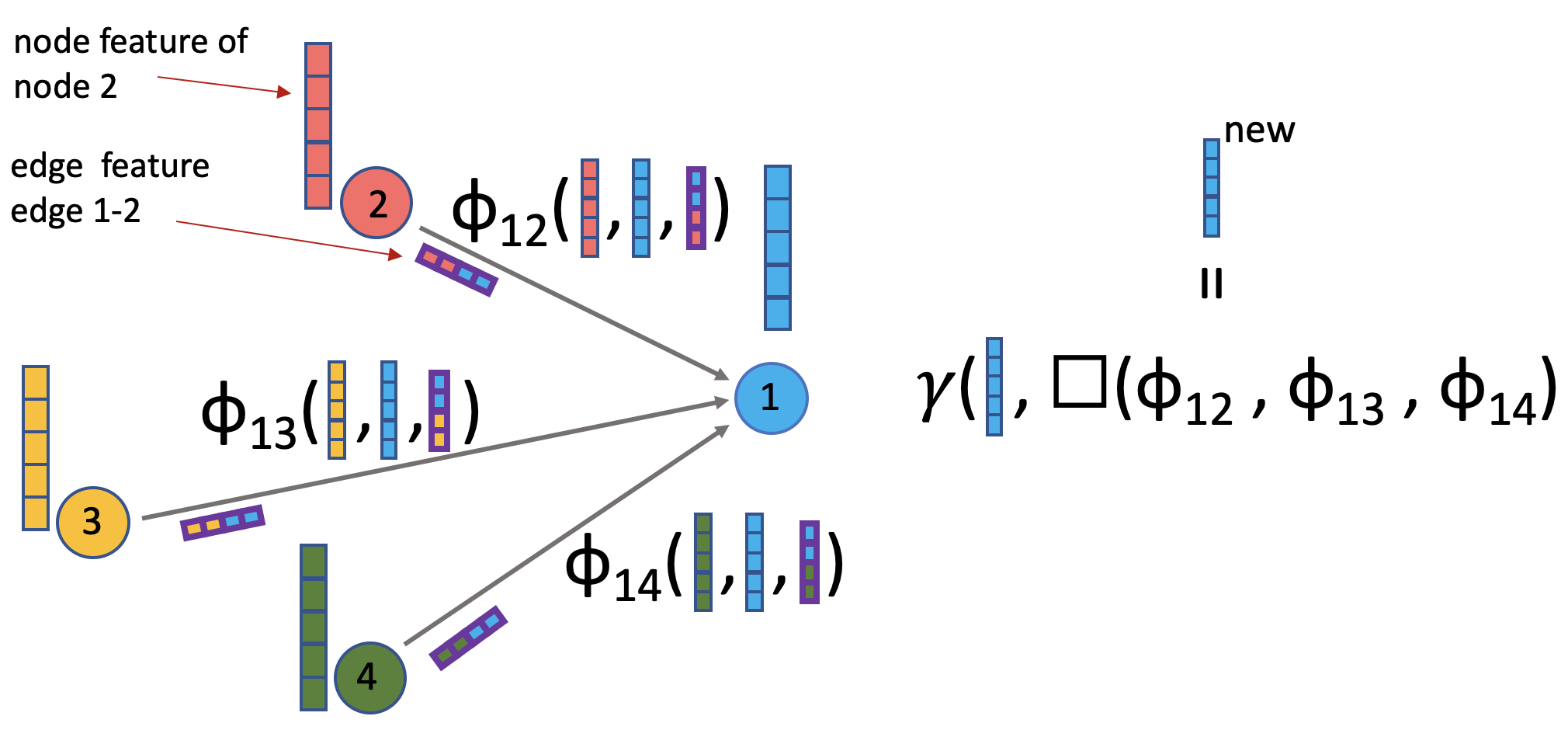}
\caption{A depiction of how message passing works in a graph neural network.}
\label{}
\end{figure}

For the GNN considered in this work our node feature vector consists of 65 elements.
\begin{enumerate}
    \item Atomic symbol (as a one hot encoded vector of [Ag, Al, As, B, Br, C, Ca, Cd, Cl, Cu, F, Fe, Ge, H , Hg , I, K, Li, Mg, Mn, N, Na, O, P, Pb, Pt, S, Se, Si, Sn, Sr, Tl , Zn, Unknown] 
    \item Degree of the atom (as a one hot encoded vector of [0, 1, 2, 3, 4, 5, 6, 7, 8, 9, 10])
    \item Implicit valence of the atom (as a one hot encoded vector of [0, 1, 2, 3, 4, 5, 6])
    \item Formal charge
    \item Number of radical electrons
    \item Hybridization of the atom (as a one hot encoded vector of [SP, SP2, SP3, SP3D, SP3D2])
    \item Is the atom aromatic? (Boolean value)
    \item Total number of hydrogen atoms (as a one hot encoded vector of [0, 1, 2, 3, 4])
\end{enumerate}
\clearpage
\section{Binning solubilities for stratified splitting of the database into train/test and validation folds}
We sample our train and test sets using a stratified sampling approach based on solubility values.
\begin{figure}[htbp]
\centering
\includegraphics[width=0.9\textwidth]{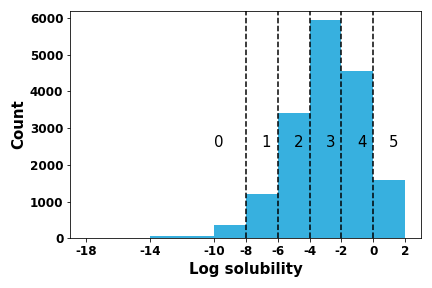}
\caption{Binned solubility distribution}
\label{}
\end{figure}

\section{Hyper-parameter tuning}
The parameters of SMILES, MDM and GNN were tuned using a combination of manual exploration and the hyperopt package~\cite{hyperopt}. 
For MDM and GNN models, 2000 random configurations were explored in order to find the best model. For the SMILES model, 35 configurations were considered. As SchNet is computationally expensive to train, we only changed the number of interaction layers from 6 to 12. We show the hyper-parameters and corresponding values explored and selected in Table S5.  We show the final architectures of the MDM and GNN models in Figure S11.

\begin{table}[!htb]
\begin{tabular}{llll}
\toprule
\textbf{Model} & \textbf{Hyper-parameter} &     \textbf{Values} & \textbf{Selected} \\
\midrule
MDM & Neurons in fully connected layers & 64 to 640 by 64 & \\
& Dropout & uniform distribution 0-1  & \\
& Activation &  relu, selu, sigmoid  & \\
& Learning rate & $10^{-3}$,$10^{-2}$, $10^{-1}$ & \\
& Optimization & adam, rmsprop, sgd & \\
& Number of fully connected layers & 2,3,4,5 & \\
\\
GNN & Node features graph layers size &  64 to 640 by 64 &  \\
& Neurons in fully connected layers & 32 to 320 by 32 &  \\
& Dropout & uniform distribution 0-1  & \\
& Activation &  relu, selu, sigmoid  & \\
& Learning rate & $10^{-3}$,$10^{-2}$, $10^{-1}$ & \\
& Optimization & adam, rmsprop, sgd & \\
& Number of GCN layers & 2,3,4 & \\
& Number of fully connected layers & 2,3,4 & \\
\\
SMI & Embedding dimension &  64 to 1088 by 64 &  \\
& Number of LSTM output units & 64 to 576 by 64  & \\
& Neurons in fully connected layers & 64 to 1088 by 64 &  \\
& Dropout & uniform distribution 0-1  & \\
& Activation &  relu, selu, sigmoid  & \\
& Learning rate & $10^{-3}$,$10^{-2}$, $10^{-1}$ & \\
& Optimization & adam, rmsprop, sgd & \\
& Number of fully connected layers & 2,3,4,5 & \\
\\
SCH & Embedding size for atoms &  64 and 128  & \\
& Number of filters & 64 and 128 &  \\
& Number of interactions & 3 to 10  & \\
& Number of layers in MLP  & 1 to 4  & \\
& Aggregation mode  & ``sum'' and ``avg''  & \\
\bottomrule
\end{tabular}
\caption{Hyper-parameters tuned and values explored for each model type}
\label{}
\end{table}

\begin{figure*}[!htb]
    \centering
    \begin{subfigure}[b]{.5\textwidth}
        \includegraphics[width=\textwidth]{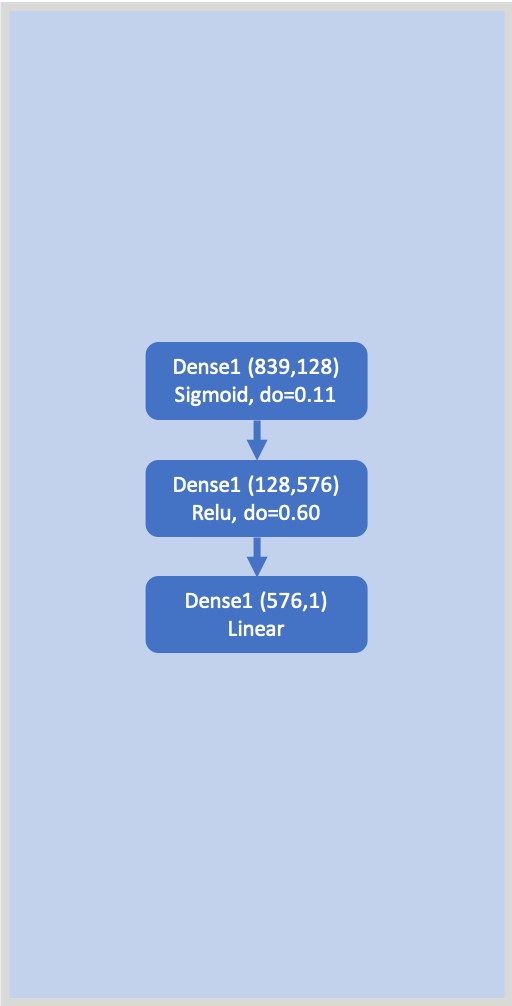}
        \caption{MDM}
        \label{}
    \end{subfigure}
    \begin{subfigure}[b]{.5\textwidth}
        \includegraphics[width=\textwidth]{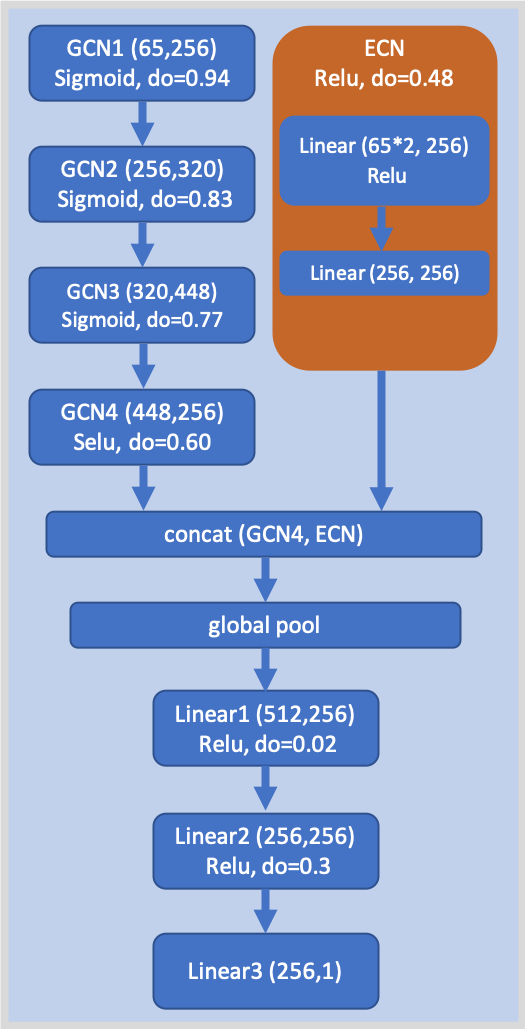}
        \caption{GNN}
        \label{}
    \end{subfigure}
    
    \caption{Final architectures of (a) MDM (implemented using Keras), and (b) GNN (implemented using PyTorch and PyTorch geometric) models. Input and output dimensions of neural network layers are given inside parenthesis respectively. ``do" stands for dropout rate.}
\end{figure*}
\clearpage

\section{Molecular Fragments Analysis}
A comparison of average errors when  molecules consisting of 1-4 fragments is tabulated in Table S6. Here, the error is defined as $|Actual - Predicted|$ solubility. In general, MDM makes better predictions for fragmented molecules than the GNN mode.

\begin{table}[!htb]
\begin{center}
\begin{tabular}{l|r|r}
\toprule
 fragments &  GNN error &  MDM error \\
\midrule
    1 &   0.717437 &   0.694818 \\
    2 &   0.652585 &   0.569263 \\
    3 &   0.846883 &   0.744782 \\
    4 &   1.469782 &   1.538846 \\
    >1 &   0.718044 &   0.632401 \\
\bottomrule
\end{tabular}
\end{center}
\caption{ GNN and MDM errors for fragmented molecules in the test set. }
\label{tbl:err-by-type}
\end{table}

\section{Cluster Analysis}

\begin{figure}[t]
\includegraphics[width=1.15\textwidth]{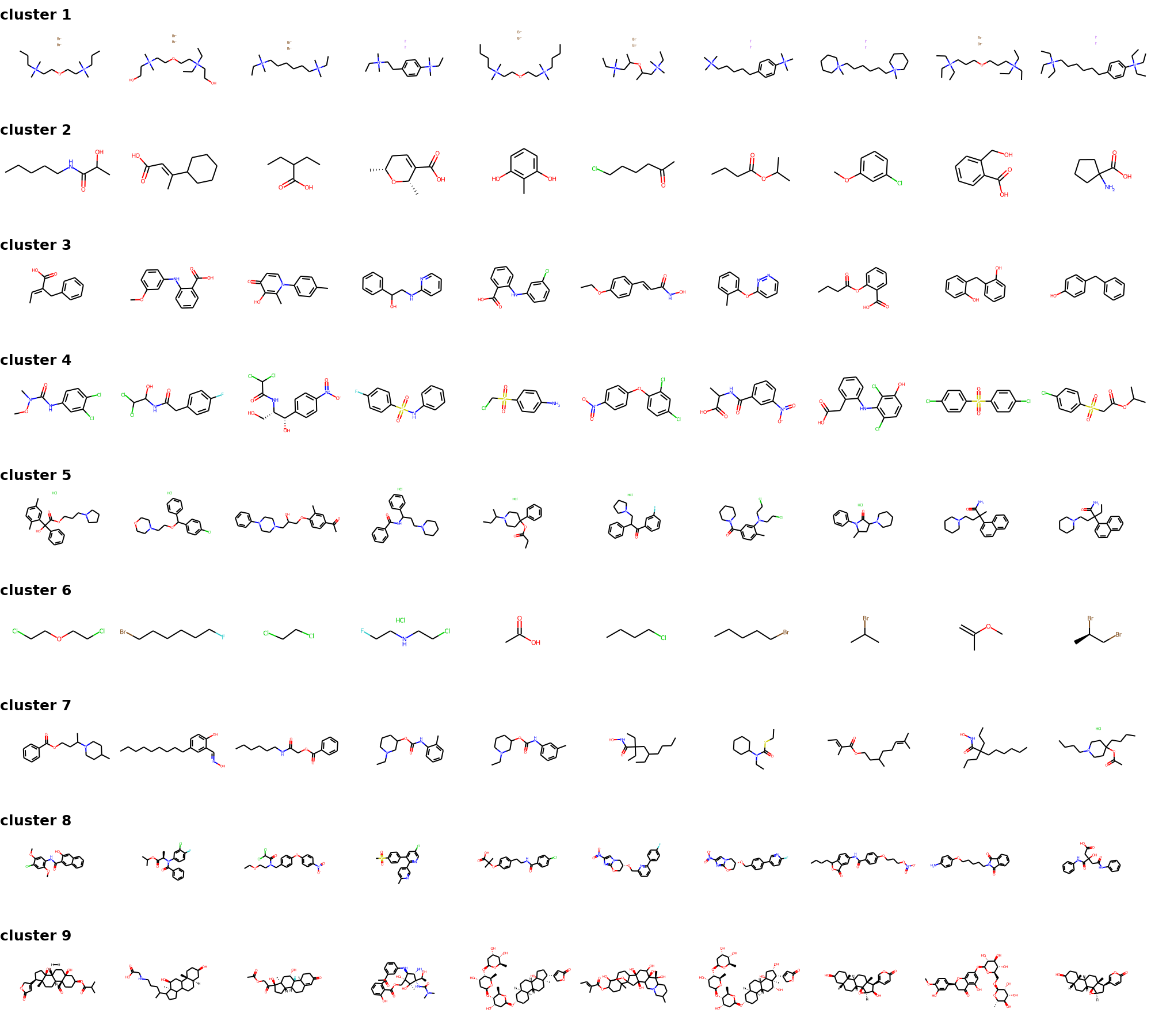}
\caption{Ten molecules closest to the 9 cluster centers.}

\label{}
\end{figure}

\begin{figure}[t]
\includegraphics[width=1.15\textwidth]{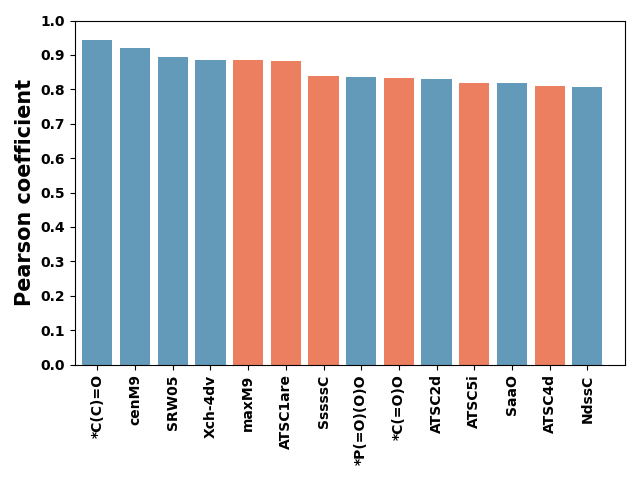}
\caption{Highly correlated descriptors with average error corresponding to clusters as discussed in section 5.1.4 of the main text}

\label{}
\end{figure}

\begin{figure}[t]
\includegraphics[width=1.15\textwidth]{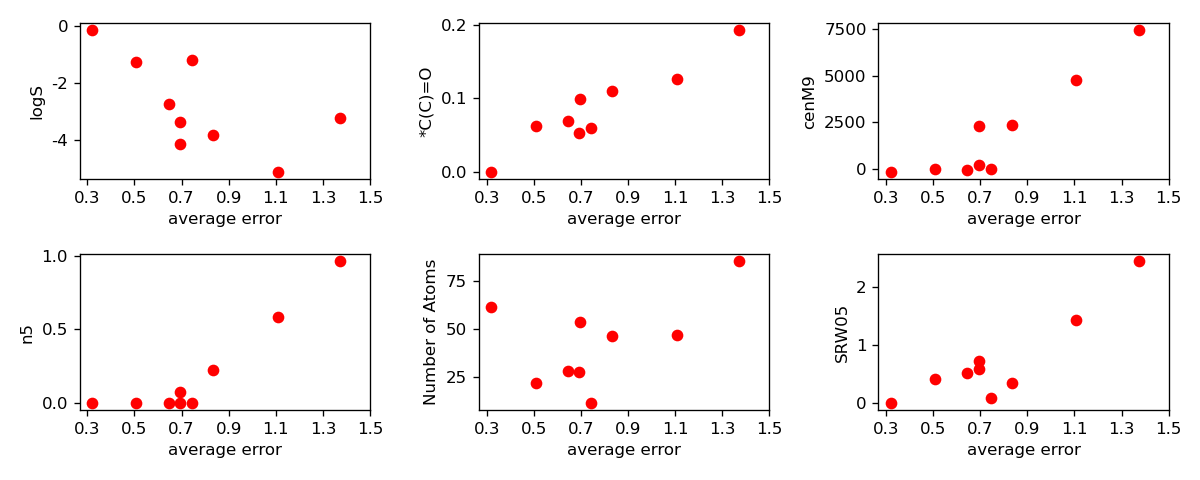}
\caption{Scatter plot of the mean value of the descriptors versus the cluster error (Section 5.1.4 in the main text).}

\label{}
\end{figure}

\begin{figure}[t]
\includegraphics[width=1.15\textwidth]{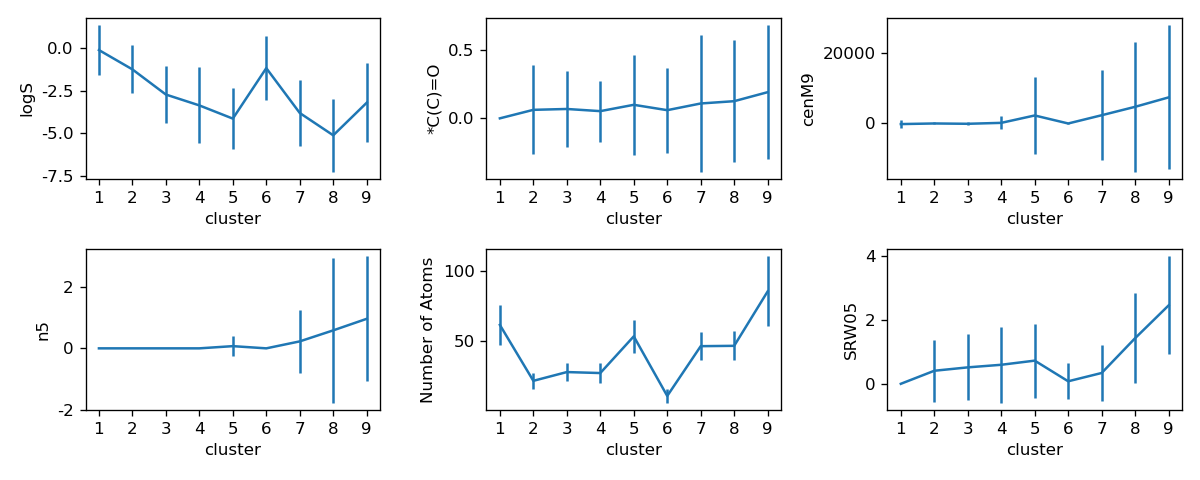}
\caption{Scatter plot of the mean value of the descriptors versus the cluster label (Section 5.1.4 in the main text).}

\label{}
\end{figure}

\clearpage
\bibliographystyle{unsrt}
\bibliography{references}